\pdfoutput=1
\documentclass[a4paper,11pt]{article}
\usepackage{jcappub} 

\usepackage[utf8]{inputenc}
\usepackage{latexsym,array,theorem,mathrsfs,bm,float}
\usepackage{psfrag}
\usepackage{amsfonts,amsmath,amssymb,latexsym,array,afterpage,theorem,mathrsfs,bm,float,epsfig,color,graphicx,tabularx,here,multirow}

\newcommand{\nn}{\nonumber \\}

\newcommand{\nablaA}{\overset{\scriptscriptstyle  \Gamma}{\nabla } {} }

\newcommand{\RA}{\overset{\scriptscriptstyle  \Gamma}{R } {} }
\newcommand{\GA}{\overset{\scriptscriptstyle  \Gamma}{G } {} }

\newcommand{\Ta}{\overset{\scriptscriptstyle  (1)}{T } {} }
\newcommand{\Tb}{\overset{\scriptscriptstyle  (2)}{T } {} }
\newcommand{\Tc}{\overset{\scriptscriptstyle  (3)}{T } {} }

\newcommand{\Ra}{\overset{\scriptscriptstyle  (1)}{R } {} }
\newcommand{\Rb}{\overset{\scriptscriptstyle  (2)}{R } {} }
\newcommand{\Rc}{\overset{\scriptscriptstyle  (3)}{R } {} }
\newcommand{\Rd}{\overset{\scriptscriptstyle  (4)}{R } {} }
\newcommand{\Ree}{\overset{\scriptscriptstyle  (5)}{R } {} }
\newcommand{\Rf}{\overset{\scriptscriptstyle  (6)}{R } {} }

\newcommand{\CA}{\overset{\scriptscriptstyle  \Gamma}{C } {} }

\newcommand{\Tcal}{ \mathcal{T} }

\newcommand{\Xcal}{ \mathcal{X} }
\newcommand{\Xs}{ {}^{\star}\! \mathcal{X} }

\newcommand{\FT}{ F }

\begin{document}
\baselineskip=12pt

\title{Consistent inflationary cosmology from quadratic gravity with dynamical torsion}
\author[a]{Katsuki Aoki,}
\emailAdd{katsuki.aoki@yukawa.kyoto-u.ac.jp}
\affiliation[a]{Center for Gravitational Physics, Yukawa Institute for Theoretical Physics, Kyoto University, 606-8502, Kyoto, Japan}

\author[a,b]{Shinji Mukohyama}
\emailAdd{shinji.mukohyama@yukawa.kyoto-u.ac.jp}
\affiliation[b]{Kavli Institute for the Physics and Mathematics of the Universe (WPI), The University of Tokyo, 277-8583, Chiba, Japan}

\date{\today}

\abstract{
The idea of gauge theories of gravity predicts that there should exist not only the massless graviton but also massive particles carrying the gravitational force. We study the cosmology in a quadratic gravity with dynamical torsion where gravity may be interpreted as a gauge force associated with the Poincar\'{e} group. In addition to the massless spin-2 graviton, the model contains four non-ghost massive particle species: a couple of spin-0, a spin-1 and a spin-2. Supposing the restoration of the local Weyl invariance in the UV limit and the parity invariance, we find the most general minisuperspace action describing a homogeneous and isotropic universe with a flat spatial geometry. We then transform the minisuperspace action to a quasi-Einstein frame in which the field space is a hyperboloid and the field potential is a combination of those of a Starobinsky-like inflation and a natural inflation. Remarkably, thanks to the multi-field dynamics, the Starobinsky-like inflationary trajectory can be realized even if the initial condition is away from the top of the Starobinsky-like potential. We also study linear tensor perturbations and find qualitatively different features than the Starobinsky inflation, spontaneous parity violation and mixing of the massless and massive spin-2 modes, which might reveal the underlying nature of gravity through inflationary observables. }

{\baselineskip0pt
\rightline{\baselineskip16pt\rm\vbox to-20pt{
           \hbox{YITP-20-23, IPMU20-0019}
\vss}}%
}
\maketitle

\section{Introduction}
Nowadays inflation is almost a part of the standard cosmology. The postulated accelerating expansion in the early epoch of the universe is not only providing a solution to some initial value problems of the Big-Bang cosmology but also explaining the origin of the primordial fluctuations. Since the fluctuations are observed precisely~\cite{Akrami:2018odb}, the inflationary cosmology can be used to test physics at the very high energy scale as high as $O(H_{\rm inf}^2M_{\rm pl}^2) =(10^{16}~{\rm GeV})^4$. At the leading order, the physics during inflation would be determined by the physics of a light particle causing the accelerating expansion, namely inflaton. From the viewpoint of the current observations, Starobinsky inflation~\cite{Starobinsky:1980te} is one of the best candidates of inflationary cosmology. On the other hand, the cosmological collider physics named by~\cite{Arkani-Hamed:2015bza} enables us to search new heavy particles with masses comparable to the inflationary Hubble scale through primordial non-Gaussianities.

Although gravity is believed to be mediated by a massless spin-2 particle, one may guess that other massive particles may contribute to additional gravitational forces at short distances just like massive weak bosons mediate the weak force. These massive particles are naturally expected to appear when the gravitational force is interpreted as a gauge force associated with the Poincar\'{e} group or the general linear group. This idea has a long history ever since the pioneering studies by Utiyama~\cite{Utiyama:1956sy}, Kibble~\cite{Kibble:1961ba} and Sciama~\cite{sciama1962analogy}; for reviews, see, e.g.,~\cite{Blagojevic:2002du,Percacci:2009ij,Blagojevic:2013xpa,Obukhov:2018bmf}. A consequence of this idea is that the spin connection (or the connection) is promoted to an object independent of the vielbein (or the metric). The underlying spacetime geometry is no longer Riemannian and the resultant theories of gravity are referred to as gauge theories of gravity or Palatini formalism of gravity depending on their context. The particles responsible for the deviations from the Riemannian geometry, described by the torsion and the non-metricity, must be massive since invariance under diffeomorphisms and local Lorentz transformations (or local general linear transformations) does not guarantee masslessness of them and thus does not protect them from acquiring non-zero masses. The additional gravitational forces are mediated by massive particles and thus cannot be seen in large distances, similarly to the weak force.

Kinetic terms of the independent connection are governed by dimension four operators such as quadratic curvature terms. There have been several attempts to clarify the particle spectrum of the gauge theories of gravity. In the present paper, we focus on theories based on the Riemann-Cartan geometry where the spacetime has the torsion as well as the curvature while the non-metricity vanishes. This could be interpreted in such a way that gravity is associated with the Poincar\'{e} group or that the non-metricity is already integrated out by supposing a sufficiently large mass hierarchy between the non-metricity and the torsion. In the context of the gauge theories of gravity, the torsion and the curvature are identified with the field strengths associated with the translation and the rotation, respectively. More attention has been paid to the so called quadratic Poincar\'{e} gauge theories (qPGTs) with the parity invariance in which the Lagrangian is schematically given by $\mathcal{L}_{\rm qPGT}=\RA+T^2+\RA^2$. Here, $\RA$ is the curvature and $T$ is the torsion with indices omitted (see, however, \cite{Karananas:2014pxa,Blagojevic:2018dpz,Aoki:2019snr,Percacci:2019hxn} for recent studies on other types of Lagrangian). It turned out that at most three massive particle species can be physical, i.e.~no ghost and no tachyon, around the Minkowski background although the generic qPGT has six massive particle species~\cite{Sezgin:1979zf,Sezgin:1981xs}. Our previous paper~\cite{Aoki:2019snr}, however, showed that one more physical particle can appear when general dimension four operators involving the derivatives of the torsion are added. We shall consider a dynamical torsion theory having massive spin-$2^+,1^+,0^+,0^-$ particle species, where the number and $\pm$ represent the spin and the parity of the particles, respectively. The massive particles with $2^+,0^+$ couple to matter via the energy-momentum tensor while the $1^+,0^-$ particles couple via the spin tensor.

It is thus interesting to search new massive particles carrying the gravitational interactions since they must trace fundamental aspects of gravity. The masses are free parameters of the theory due to the lack of complete understanding of quantum gravity and should be constrained experimentally.
In the present paper, we suppose that the masses are of order of $10^{13}$ GeV in order that the dimension four operators of the theory are responsible for the origin of the cosmic inflation. 
Inflation in the quadratic gravity with the torsion is closely related to Starobinky's inflationary model. Indeed, we will find that the prediction of the Starobinsky model is recovered under a certain limit of the parameters where the massive $2^+,1^+,0^-$ particles become infinitely heavy to be integrated out while the mass of the $0^+$ particle is kept finite. Nevertheless, the underlying nature of gravity is different in these models and the difference may be traced by (non-)existence of other massive particle species\footnote{Recently, studies on inflation in the Palatini formalism of gravity has gained increasing attention (see \cite{Tenkanen:2020dge} for a review). In this context, it is usually supposed that the torsion vanishes whereas the non-metricity does not. Rather than this point, it would be worth emphasizing a difference from the present study: these studies have focused on Lagrangians in which the non-metricity (and the torsion) is infinitely massive and assumed that inflation is caused by another scalar field such as Higgs field. For instance, the Lagrangian \eqref{L_light} generally has the massive spin-0 mode but its mass becomes infinite when $\alpha_{T_2}=1$. One needs to put another field to realize inflation in this case.}.
As a first step, the present paper is devoted to a study of the background dynamics of the universe in a generic parameter space of the theory and briefly discuss the linear tensor perturbations. Even so, we will find that the existence of additional particles, especially $2^+$ and $0^-$, can yield qualitatively different features than the Starobinsky model.

The rest of the present paper is organized as follows. We first summarize the notation and formulate the Lagrangian in Sec.~\ref{sec_QG}. Sec.~\ref{sec_heavy} is devoted to the case where the $2^+,1^+,0^-$ particles are infinitely heavy and integrated out. The relation between the Starobinsky model and the generic quadratic gravity with the dynamical torsion is clarified. We then study the background dynamics of the universe in the generic parameter space of the theory in Sec.~\ref{sec_BG}. We discuss tensor perturbations in Sec.~\ref{sec_tensor} and  finally make summary remarks in Sec.~\ref{summary}.

\section{Quadratic gravity with dynamical torsion}
\label{sec_QG}
\subsection{Notation}
In the present paper, we study the cosmology in a parity invariant gravitational theory where the torsion is supposed to be dynamical. The signature of the metric is $(-,+,+,+)$. The tetrad and the spin connection are commonly used to describe gravitational theories with a dynamical torsion.
We denote the most general form of the action as
\begin{align}
\mathcal{L}(e^a_{\mu},\omega^{ab}{}_{\mu})=\frac{M_{\rm pl}^2}{2}\mathcal{L}_2+\frac{M_{\rm pl}^2}{M_*^2} \mathcal{L}_4+\sum_{n>4}\frac{M_{\rm pl}^2}{\Lambda^{n-2}}\mathcal{L}_n\,,
\label{EFT_action}
\end{align}
where $\mathcal{L}_2$ and $\mathcal{L}_4$ are operators of scaling dimension two and four, respectively, and $\mathcal{L}_n$ are terms involving higher dimensional operators. The tetrad and the spin connection are supposed to be independent variables of which scaling dimensions are $[e^a_{\mu}]=0$ and $[\omega^{ab}{}_{\mu}]=1$, respectively. 
Greek indices $\mu,\nu,\cdots$ are used to denote spacetime indices whereas Latin indices $a,b,\cdots$ are used to represent the Lorentz indices.
We assume the zero cosmological constant to admit the Minkowski vacuum. The dimension four operators are factorized by $M_{\rm pl}^2/M_*^2$ for convenience. We throughout assume $M_*/M_{\rm pl} \ll 1$ which may be interpreted as a weak coupling of the connection since the dimension four operators lead to kinetic terms of the independent connection. 
For simplicity, we consider the metric compatible connection, i.e. the spin connection has the antisymmetric Lorentz indices, where the resultant geometry is called the Riemann-Cartan geometry. Due to the general covariance and the local Lorentz invariance, the action should be built out of the geometrical quantities and their covariant derivatives where the Riemann-Cartan curvature and the torsion tensor are defined by
\begin{align}
\RA^{ab}{}_{\mu\nu}&:=2\partial_{[\mu}\omega^{ab}{}_{\nu]}+2\omega^a{}_{c[\mu}\omega^{cb}{}_{\nu]}
\,, \\
T^a{}_{\mu\nu}&:=2\partial_{[\mu}e^{a}_{\nu]}+2\omega^{a}{}_{b[\mu}e^{b}_{\nu]}
\,.
\end{align}
The covariant derivative is denoted by $\nablaA_{\mu}$ in the Riemann-Cartan geometry. Since the scaling dimension are $[\RA^{ab}{}_{\mu\nu}]=2$ and $[T^a{}_{\mu\nu}]=1$, admitting the equivalence upon the integration by parts, the scaling dimension four operator $\mathcal{L}_4$ may not contain the covariant derivatives of the curvature, namely
\begin{align}
\mathcal{L}_4=\mathcal{L}_4(\eta_{ab},e^a_{\mu},T^{a}{}_{\mu\nu},\RA^{ab}{}_{\mu\nu},\nablaA_{\mu}T^a{}_{\nu\rho})\,.
\end{align}
The generic form of the dimension four terms with the parity invariance was studied in \cite{christensen1980second,Aoki:2019snr}.

On the other hand, we will use the coordinate basis expressions in which the local Lorentz invariance becomes manifest. The metric $g_{\mu\nu}$ and the connection $\Gamma^{\mu}_{\nu\rho}$ are related to the tetrad and the spin connection via
\begin{align}
g_{\mu\nu}&=\eta_{ab}e^{a}_{\mu}e^b_{\nu} \,, \\
\nablaA_{\mu}e^{a}_{\nu}&=\partial_{\mu}e^{a}_{\nu}+\omega^{a}{}_{b\mu}e^b_{\nu}-\Gamma^{\rho}_{\nu\mu}e^a_{\rho}=0
\,.
\end{align}
The Riemann-Cartan curvature and the torsion are given by
\begin{align}
\RA{}^{ \mu}{}_{\nu\alpha\beta}(\Gamma)&=2\partial_{[\alpha}\Gamma^{\mu}_{\nu|\beta]}
+\Gamma^{\mu}_{\sigma[\alpha } \Gamma^{\sigma}_{\nu|\beta] }\,,
\\
T^{\mu}{}_{\alpha\beta}&=2\Gamma^{\mu}_{[\beta\alpha]}\,,
\end{align}
in the coordinate basis. The Ricci tensor and the Ricci scalar are defined by
\begin{align}
\RA_{\mu\nu}:=\RA^{\alpha}{}_{\mu\alpha\nu}\,, \quad
\RA:=g^{\mu\nu}\RA_{\mu\nu}\,,
\end{align}
and the Einstein tensor is
\begin{align}
\GA_{\mu\nu} &:=\RA_{\mu\nu}-\frac{1}{2}g_{\mu\nu}\RA
\,.
\end{align}
Note that the metric and the connection are not independent variables due to the metric compatibility condition
\begin{align}
\nablaA_{\mu}g^{\nu\rho}=0\,.
\end{align}
Instead, one can choose the metric and the torsion as independent variables. The metric-compatible connection $\Gamma^{\mu}_{\alpha\beta}$ is then computed as
\begin{align}
\Gamma^{\mu}_{\alpha\beta}=
\left\{ {}^{\,\, \mu}_{\alpha\beta} \right\}
-\frac{1}{2}(T^{\mu}{}_{\alpha\beta}-T_{\beta}{}^{\mu}{}_{\alpha}+T_{\alpha\beta}{}^{\mu} )
\,, \label{connection}
\end{align}
where $\left\{ {}^{\,\, \mu}_{\alpha\beta} \right\}$ is the Levi-Civita connection. We use $R^{\mu}{}_{\nu\alpha\beta}$ and $\nabla_{\mu}$ to refer the Riemann curvature and the covariant derivatives in the Riemannian geometry, i.e. $R^{\mu}{}_{\nu\alpha\beta}$ and $\nabla_{\mu}$ are defined by the Levi-Civita connection.

For later convenience, we decompose the Riemann-Cartan curvature and the torsion into irreducible pieces. As for the Riemann curvature $R^{\mu}{}_{\nu\alpha\beta}$, there are three irreducible pieces, namely, the Weyl tensor, the traceless part of the Ricci tensor, and the Ricci scalar. On the other hand, the Riemann-Cartan curvature are decomposed into six pieces (see e.g.~\cite{Obukhov:2018bmf,Blagojevic:2018dpz}) where three of them are the counterparts of the three irreducible pieces of $R^{\mu}{}_{\nu\alpha\beta}$ while the other three are obtained from the non-Riemannian part of the curvature
\begin{align}
\Xcal^{\mu\nu}:=\frac{1}{2}\RA^{\mu}{}_{\alpha\beta\gamma}\epsilon^{\alpha\beta\gamma\nu}\,.
\end{align}
The irreducible pieces are given by
\begin{align}
\Rb_{\mu\nu\rho\sigma}&:=\frac{1}{2}(g_{\mu\alpha}\Xcal^T_{\nu\beta}-g_{\nu\alpha}\Xcal^T_{\mu\beta})\epsilon^{\alpha\beta}{}_{\rho\sigma}
\,, \\
\Rc_{\mu\nu\rho\sigma}&:=\frac{1}{12}\Xcal \epsilon_{\mu\nu\rho\sigma}
\,, \\
\Rd_{\mu\nu\rho\sigma}&:=g_{\mu[\rho}\RA^T_{\nu|\sigma]}-g_{\nu[\rho}\RA^T_{\mu|\sigma]}
\,, \\
\Ree_{\mu\nu\rho\sigma}&:=g_{\mu[\rho}{}^{\star}\!\Xcal_{\nu|\sigma]}-g_{\nu[\rho}{}^{\star}\!\Xcal_{\mu|\sigma]}
\,, \\
\Rf_{\mu\nu\rho\sigma}&:=\frac{1}{6}\RA g_{\mu[\rho}g_{\nu|\sigma]}
\,, \\
\Ra_{\mu\nu\rho\sigma}&:=\RA_{\mu\nu\rho\sigma}-\sum_{n=2}^6 \overset{\scriptscriptstyle  (n)}{R } {}_{\mu\nu\rho\sigma}\,,
\end{align}
where
\begin{align}
\Xcal&:=\Xcal^{\mu}{}_{\mu}=-\frac{1}{2}\epsilon^{\mu\nu\rho\sigma}\RA_{\mu\nu\rho\sigma}
\,, \\ 
\Xcal^T_{\mu\nu}&:=\Xcal_{(\mu\nu)}-\frac{1}{4}g_{\mu\nu} \Xcal 
\,, \\
{}^{\star}\!\Xcal_{\mu\nu}&:=\frac{1}{2}\epsilon_{\mu\nu\alpha\beta}\Xcal^{[\alpha\beta]}=\RA_{[\mu\nu]}
\,, \\
\RA^T_{\mu\nu}&:=\RA_{(\mu\nu)}-\frac{1}{4}g_{\mu\nu} \RA\,.
\end{align}
The irreducible piece $\Ra_{\mu\nu\rho\sigma}$ satisfies the same symmetric properties of the Weyl tensor,
\begin{align}
\Ra_{\mu\nu(\rho\sigma)}=\Ra_{(\mu\nu)\rho\sigma}=0
\,,\quad \Ra_{\mu\nu\rho\sigma}=\Ra_{\rho\sigma\mu\nu}
\,, \quad \Ra_{\mu[\nu\rho\sigma]}=0
\,, \quad \Ra^{\mu}{}_{\nu\mu\rho}=0\,.
\end{align}
On the other hand, the torsion tensor is decomposed into three irreducible pieces,
\begin{align}
\Tb_{\mu\nu\rho} &=\frac{2}{3}g_{\mu[\nu}T_{\rho]}
, \\
\Tc_{\mu\nu\rho}&=\epsilon_{\mu\nu\rho\sigma}\mathcal{T}^{\sigma} \label{T3}
, \\
\Ta_{\mu\nu\rho}&=T_{\mu\nu\rho}-\Tb_{\mu\nu\rho}-\Tc_{\mu\nu\rho}\,,
\end{align}
with
\begin{align}
T_{\mu}&:=T^{\nu}{}_{\nu\mu}\,, \\
\mathcal{T}_{\mu}&:=\frac{1}{6}\epsilon_{\mu\nu\rho\sigma}T^{\nu\rho\sigma}\,.
\end{align}
The irreducible piece $\Ta_{\mu\nu\rho}$ satisfies the following identities
\begin{align}
\Ta_{\mu(\nu\rho)}=0\,,\quad \Ta_{[\mu\nu\rho]}=0\,, \quad \Ta^{\mu}{}_{\mu\nu}=0\,.
\end{align}

For convenience, we also define the traceless part of the Riemann-Cartan curvature,
\begin{align}
\CA_{\mu\nu\rho\sigma}:\!&
=\Ra_{\mu\nu\rho\sigma}+\Rb_{\mu\nu\rho\sigma}+\Rc_{\mu\nu\rho\sigma}
\nn
&=\RA_{\mu\nu\rho\sigma}-\left(g_{\mu[\rho}\RA_{\nu|\sigma]}-g_{\nu[\rho}\RA_{\mu|\sigma]}\right)+\frac{1}{3}g_{\mu[\rho}g_{\nu|\sigma}\RA
\,,
\end{align}
and its square
\begin{align}
\CA^2&:=-\frac{1}{4}\epsilon^{\mu\nu\rho\sigma}\epsilon^{\alpha\beta\gamma\delta}\CA_{\mu\nu\alpha\beta}\CA_{\rho\sigma\gamma\delta}
\nn
&\,\,=\CA_{\mu\nu\rho\sigma}\CA^{\rho\sigma\mu\nu}
=\Ra_{\mu\nu\rho\sigma}\Ra^{\mu\nu\rho\sigma}+2\Xcal^T_{\mu\nu}\Xcal^{T\mu\nu}-\frac{1}{6}\Xcal^2 \,.
\end{align}
Although the expression of $\CA_{\mu\nu\rho\sigma}$ is similar to the Weyl tensor in the Riemannian geometry, $\CA_{\mu\nu\rho\sigma}$ does not satisfy the symmetric properties of the Weyl tensor. The irreducible piece is not $\CA_{\mu\nu\rho\sigma}$ but $\Ra_{\mu\nu\rho\sigma}$.
We also denote the Gauss-Bonnet term as
\begin{align}
\RA^2_{\rm GB}:\!&
=-\frac{1}{4}\epsilon^{\mu\nu\rho\sigma}\epsilon^{\alpha\beta\gamma\delta}\RA_{\mu\nu\alpha\beta}\RA_{\rho\sigma\gamma\delta}
\nn
&=\Ra_{\mu\nu\rho\sigma}\Ra^{\mu\nu\rho\sigma}
-\Rb_{\mu\nu\rho\sigma}\Rb^{\mu\nu\rho\sigma}
+\Rc_{\mu\nu\rho\sigma}\Rc^{\mu\nu\rho\sigma}
-\Rd_{\mu\nu\rho\sigma}\Rd^{\mu\nu\rho\sigma}
+\Ree_{\mu\nu\rho\sigma}\Ree^{\mu\nu\rho\sigma}
+\Rf_{\mu\nu\rho\sigma}\Rf^{\mu\nu\rho\sigma}
\nn
&=\RA^{\mu\nu\rho\sigma}\RA_{\rho\sigma\mu\nu}-4\RA^{\mu\nu}\RA_{\nu\mu}+\RA^2\,,
\end{align}
which is a boundary term in four dimensions.

\subsection{Lagrangian with asymptotic Weyl invariance}
Among general possibilities of Lagrangian, we only consider a Lagrangian having a certain property. A special class, which has gained attention in the literature, is called the quadratic Poincar\'{e} gauge theories (qPGTs) where the dimension four terms are supposed to be quadratic in the Riemann-Cartan curvature:
\begin{align}
\mathcal{L}_{{\rm qPGT},4}&=\sum_{n=1}^6 b_n \overset{\scriptscriptstyle  (n)}{R } {}_{\mu\nu\rho\sigma}\overset{\scriptscriptstyle  (n)}{R } {}^{\mu\nu\rho\sigma} 
\,, \label{PGT4}
\end{align}
where $b_n$ are dimensionless constants.
The restriction to only the quadratic curvature terms must be spoiled by radiative corrections unless protected by some mechanism, which may motivate us to generalize the qPGTs. The quadratic curvature terms \eqref{PGT4} have a property different from other generic dimension four terms: the corresponding action of \eqref{PGT4} is invariant under the local Weyl transformation in the sense of the Riemann-Cartan geometry~\cite{Obukhov:1982zn},
\begin{align}
e^a_{\mu} \rightarrow e^{\Omega(x)}e^a_{\mu} \,, \quad \omega^{ab}{}_{\mu} \rightarrow \omega^{ab}{}_{\mu} \,.
\label{Weyl_trans}
\end{align}
Although the local Weyl invariance cannot be an exact symmetry of the theory in order to include dimension two terms especially the Einstein-Hilbert action, the asymptotic restoration of the certain symmetry in the UV limit could be used to constrain the possible terms of the Lagrangian. 
We thus assume throughout the present paper the ``asymptotic'' local Weyl invariance where the dimension four operators respect the local Weyl invariance \eqref{Weyl_trans} while the lower dimensional operators $\mathcal{L}_2$ do not. This situation may be realized if the renormalization group (RG) flow of the underlining theory admits a UV fixed point with the local Weyl invariance. The higher dimensional operators $\mathcal{L}_n~(n>4)$ may be included but in the present paper we assume the hierarchy $M_* \ll \Lambda$ so that they can be ignored for the study of inflationary dynamics at energy scales up to $M_*$. This assumption may be justified if the RG flow admits a saddle point with the local Weyl invariance and the flow stays near the saddle point for a sufficiently long at intermediate scales $M_*\lesssim E \ll \Lambda$.

One can introduce a dilaton field, coupling to the dimension two operators properly, as a St\"ueckelberg field to restore the invariance under the local Weyl transformation at all scales. After introducing the dilaton, the asymptotic local Weyl invariance would be reinterpreted as the assumption that the dilaton is decoupled in the UV limit. However, in the present paper we shall work on the ``unitary gauge'' Lagrangian where the dilaton takes a fixed value and does not appear explicitly.

The Weyl transformation \eqref{Weyl_trans} is equivalent to the Weyl rescaling of the metric accompanied by the integrable projective transformation
\begin{align}
g_{\mu\nu}\rightarrow e^{2\Omega(x)}g_{\mu\nu} \,, \quad 
\Gamma^{\mu}_{\alpha\beta} \rightarrow \Gamma^{\mu}_{\alpha\beta}+\delta^{\mu}_{\alpha}\partial_{\beta}\Omega \,,
\label{Weyl_metric}
\end{align}
under which the Riemann-Cartan curvature $\RA^{\mu}{}_{\nu\alpha\beta}$ is invariant while the torsion is transformed as
\begin{align}
T^{\mu}{}_{\nu\rho} \rightarrow T^{\mu}{}_{\nu\rho}-2 \delta^{\mu}_{[\nu}\partial_{\rho]}\Omega 
\,.
\end{align}
The transformation law of the torsion implies that only the irreducible pieces $T_{\mu}$ is transformed under \eqref{Weyl_metric} and the others are invariant:
\begin{align}
T_{\mu} \rightarrow T_{\mu}-3\partial_{\mu} \Omega\,, \quad 
\Ta^{\mu}{}_{\nu\rho} \rightarrow \Ta^{\mu}{}_{\nu\rho}\,, \quad \Tcal_{\mu} \rightarrow \Tcal_{\mu} \,.
\label{Weyl_torsion}
\end{align}

Since the transformation law of $T_{\mu}$ is the same as that of the $U(1)$ gauge transformation, we obtain an invariant tensor,
\begin{align}
\FT_{\mu\nu}:=2\partial_{[\mu}T_{\nu]}\,,
\end{align}
composed of the derivative of $T_{\mu}$. We can also consider Weyl invariant tensors made by derivatives of the other irreducible pieces $\Ta^{\mu}{}_{\nu\rho}$ and $\Tcal_{\mu}$. The simplest way finding them is to define a new connection
\begin{align}
\tilde{\Gamma}^{\mu}_{\alpha\beta}:=\Gamma^{\mu}_{\alpha\beta}+\frac{1}{3}\delta^{\mu}_{\alpha}T_{\beta}
\,, \label{Gamma_tilde}
\end{align}
and to consider covariant derivatives with respect to \eqref{Gamma_tilde}. The new connection \eqref{Gamma_tilde} is invariant under \eqref{Weyl_metric}. Hence, the following tensors
\begin{align}
\tilde{\nabla}_{\mu}\Tcal_{\nu}\,, \quad \tilde{\nabla}_{\mu}\Ta^{\nu}{}_{\rho\sigma}\,, \label{til_nabla}
\end{align}
are also local Weyl invariant where $\tilde{\nabla}_{\mu}$ is the covariant derivative with respect to the deformed connection \eqref{Gamma_tilde}.
Note that $\Xcal^{\mu\nu}$ is written in terms of the torsion tensor by virtue of the Bianchi identity as
\begin{align}
\Xcal^{\mu\nu}&=\frac{1}{2}(\nablaA_{\alpha}T^{\mu}{}_{\beta\gamma}-T^{\mu}{}_{\alpha\delta}T^{\delta}{}_{\beta\gamma})\epsilon^{\alpha\beta\gamma\nu} 
\nn
&=(g^{\mu\nu}g^{\alpha\beta}-g^{\mu\alpha}g^{\nu\beta})\tilde{\nabla}_{\alpha}\Tcal_{\beta}-\frac{1}{6}\FT_{\alpha\beta}\epsilon^{\alpha\beta\mu\nu}+\frac{1}{2}(\tilde{\nabla}_{\alpha}\Ta^{\mu}{}_{\beta\gamma}-\Ta^{\mu}{}_{\alpha\delta}\Ta^{\delta}{}_{\beta\gamma})\epsilon^{\alpha\beta\gamma\nu} 
+2\Ta^{(\mu\nu)\alpha}\Tcal_{\alpha}
\,,
\end{align}
and then one of $\Xcal_{\mu\nu}$ and $\tilde{\nabla}_{\mu}\Tcal_{\nu}$ are redundant. We do not need to use $\tilde{\nabla}_{\mu}\Tcal_{\nu}$ as an independent ingredient of the Lagrangian.

The tensor $\FT_{\mu\nu}$ and the deformed connection \eqref{Gamma_tilde} have a certain geometrical meaning.
Two connections $\Gamma^{\mu}_{\alpha\beta}$ and $\tilde{\Gamma}^{\mu}_{\alpha\beta}$ are related via the projective transformation. The geometry with the deformed connection \eqref{Gamma_tilde} is not metric compatible but satisfies the so-called Weyl's semi-metricity condition $\tilde{\nabla}_{\mu}g^{\alpha\beta}=\frac{2}{3}T_{\mu}g^{\alpha\beta}$ where the torsion vector $T_{\mu}$ plays a role of the Weyl vector. The curvature and the torsion of $\tilde{\Gamma}^{\mu}_{\alpha\beta}$ are given by
\begin{align}
\tilde{R}^{\mu}{}_{\nu\rho\sigma}&=\RA^{\mu}{}_{\nu\alpha\beta}+\frac{2}{3}\delta^{\mu}{}_{\nu}\FT_{\rho\sigma}
\,, \\
\tilde{T}^{\mu}{}_{\nu\rho}&=\Ta^{\mu}{}_{\nu\rho}+\Tc^{\mu}{}_{\nu\rho}
\,.
\end{align}
The resultant geometry is a Weyl-Cartan geometry with a traceless torsion, $\tilde{T}^{\mu}{}_{\mu\nu}=0$, where both curvature and torsion are now Weyl invariant. The tensor $\FT_{\mu\nu}$ can be interpreted as an irreducible component of the curvature $\tilde{R}^{\mu}{}_{\nu\rho\sigma}$ or can be interpreted as the field strength of the Weyl vector $T_{\mu}$.

As a result, the manifestly local Weyl invariant Lagrangian in the Riemann-Cartan geometry can be constructed from the following building blocks,
\begin{align}
\Ta^{\mu}{}_{\nu\rho}\,, ~
\Tcal_{\mu}\,, ~
\RA^{\mu}{}_{\nu\rho\sigma}
\,, ~
\FT_{\mu\nu}
\,, ~ 
\tilde{\nabla}_{\mu}\Ta^{\nu}{}_{\rho\sigma}
\,, \label{building_blocks}
\end{align}
where the first two tensors are of dimension one whereas others are of dimension two.
Any dimension four Lagrangian $\mathcal{L}_4$ constructed from \eqref{building_blocks} and the metric tensor is transformed as
\begin{align}
 \mathcal{L}_4 \rightarrow e^{-4\Omega} \mathcal{L}_4\,,
\end{align}
under \eqref{Weyl_trans}, and then the corresponding action $S_4=\int d^4x \sqrt{-g}\frac{M_{\rm pl}^2}{M_*^2} \mathcal{L}_4$ is local Weyl invariant.

However, we still have many possible terms with the scaling dimension four even if the local Weyl invariance is imposed. 
To simplify the Lagrangian we ignore terms involving $\tilde{\nabla}_{\mu}\Ta^{\nu}{}_{\rho\sigma}$ and consider dimension four operators of the form
\begin{align}
\mathcal{L}_4=\mathcal{L}_4(g^{\mu\nu},\Ta^{\mu}{}_{\nu\rho},\Tc^{\mu}{}_{\nu\rho},\RA^{\mu}{}_{\nu\rho\sigma},\FT_{\mu\nu})
=\mathcal{L}_4(g^{\mu\nu},\tilde{T}_{\mu\nu\rho},\tilde{R}_{\mu\nu\rho\sigma})\,,
\end{align}
where the Lagrangian may be regarded as a PGT deformed by the projective transformation. This ansatz of the Lagrangian does not lose the generality for the analysis of the background universe since the tensor $\Ta_{\mu\nu\rho}$ identically vanishes in the homogeneous and isotropic universe. On the other hand, the ignored terms contribute to perturbation around the cosmological background and thus the discussion on the tensor perturbations is not general.

\subsection{Ghost-free theory with four massive particle species}
The system described by the generic Lagrangian \eqref{LG} below, with \eqref{L_2}, \eqref{L_4^2}, \eqref{L_4^3} and \eqref{L_4^4}, contains massive $2^{\pm},1^{\pm},0^{\pm}$ particle species, where the number and $\pm$ denote the spin and the parity of the particle species, which include both non-ghost and ghost particle species. If the coupling constants appearing in $\mathcal{L}_2,\mathcal{L}_4$ are of $O(1)$ without any fine-tuning, masses of the ghosts are of order $M_*$. Since the inflationary Hubble scale is also $M_*$ as we will see latter, the theory can consistently describe the inflationary universe only if the coupling constants are fine-tuned in order that the masses of ghosts become sufficiently heavier than $M_*$. In the present paper, we simply consider a theory where the ghost masses are infinitely heavy, namely a ghost-free theory. Note that we refer to theories as ``ghost-free'' if theories have no ghost at least around the cosmological background as well as weakly curved backgrounds. Since \eqref{EFT_action} must be a low energy EFT of a fundamental theory, we do not require that ``ghost-free'' theories are free from ghosts around arbitrary backgrounds.

The dimension four operators can be classified into
\begin{align}
\mathcal{L}_4=\mathcal{L}_4^{(2)}+\mathcal{L}_4^{(3)}+\mathcal{L}_4^{(4)}\,,
\end{align}
where $\mathcal{L}_4^{(i)}$ ($i=2,3,4$) are the covariant terms starting from quadratic, cubic, and quartic orders of perturbations around the Minkowski background, respectively. 
The Lagrangian studied in the present paper is therefore
\begin{align}
\mathcal{L}_G=\frac{M_{\rm pl}^2}{2}\mathcal{L}_2+\frac{M_{\rm pl}^2}{M_*^2}\left( \mathcal{L}_4^{(2)}+\mathcal{L}_4^{(3)}+\mathcal{L}_4^{(4)}\right)\,.
\label{LG}
\end{align}
The stability conditions on the Minkowski spacetime yield constraints on $\mathcal{L}_2$ and $\mathcal{L}_4^{(2)}$. The generic forms are
\begin{align}
\mathcal{L}_2&=\RA+a_1 \Ta_{\mu\nu\rho}\Ta^{\mu\nu\rho}+a_2T_{\mu}T^{\mu}+ a_3 \Tcal_{\mu}\Tcal^{\mu}
\,,
\label{L_2}
\\
\mathcal{L}_4^{(2)}&=\sum_{n=1}^6 b_n \overset{\scriptscriptstyle  (n)}{R } {}_{\mu\nu\rho\sigma}\overset{\scriptscriptstyle  (n)}{R } {}^{\mu\nu\rho\sigma} + b_7 \FT_{\mu\nu}\FT^{\mu\nu}+b_8 \FT_{\mu\nu}\Xs^{\mu\nu}
\nn
&=b_1 \Ra^{\mu\nu\rho\sigma}\Ra_{\mu\nu\rho\sigma}-2b_2 \Xcal^T_{\mu\nu}\Xcal^{T\mu\nu}-\frac{b_3}{6}\Xcal^2+2b_4\RA^T_{\mu\nu}\RA^{T\mu\nu}+2b_5\Xs_{\mu\nu}\Xs^{\mu\nu}+\frac{b_6}{6}\RA^2
\nn
&+ b_7 \FT_{\mu\nu}\FT^{\mu\nu}+b_8 \FT_{\mu\nu}\Xs^{\mu\nu}
\,, \label{L_4^2} 
\end{align}
where $a_i$ and $b_n$ ($i=1,2,3; n=1,\cdots,8$) are dimensionless constants. The qPGT is the case with $b_7=b_8=0$. Whereas the qPGTs can have at most three massive particle species~\cite{Sezgin:1979zf,Sezgin:1981xs} in addition to the massless graviton, our previous paper showed that at most four different particle species can coexist around the Minkowski background in general theories~\cite{Aoki:2019snr}. The theory with non-ghost massive $0^+,0^-,1^+,2^+$ particle species is obtained when three critical conditions
\begin{align}
b_1+b_2=0\,, \quad 8(b_4+b_5) b_7-b_8^2&=0 \,,
\nn
3[a_2(b_4+b_5)+3b_7+b_8]+2a_1(2b_4+2b_5+9b_7+3b_8)&=0
\,,
\end{align}
are satisfied. The critical conditions eliminate ghost spin $2^-$ and $1^-$ modes around the Minkowski background. Only 8 parameters of the 11 parameters $(a_i,b_n)$ can be free parameters. A useful parametrization of the coupling constants is
\begin{align}
a_1&=\frac{1}{2}(\alpha_{T_1}-1),~a_2=-\frac{2}{3\alpha^2}(\alpha_{T_1}-\alpha^2),~a_3=\frac{3}{2}(\alpha_{T_3}-1),~
b_1=\frac{\alpha_C}{4}+\alpha_{\rm GB},
\nn
b_2&=-\frac{\alpha_C}{4}-\alpha_{\rm GB},~b_3=-\frac{\alpha_{\Xcal}}{2}+\frac{\alpha_C}{4}+\alpha_{\rm GB},~
b_4=-\alpha_{\rm GB},~ b_5=\frac{1}{4}(3\alpha_Y+\alpha_C)+\alpha_{\rm GB},
\nn
b_6&=\frac{\alpha_R}{2}+\alpha_{\rm GB},~b_7=\frac{1}{18\alpha^2}(3\alpha_Y+\alpha_C)(\alpha-1)^2,~
b_8=\frac{1}{3\alpha}(3\alpha_Y+\alpha_C)(1-\alpha),
\label{GF_para}
\end{align}
by the use of 8 parameters $(\alpha,\alpha_{T_1},\alpha_{T_3},\alpha_R,\alpha_C,\alpha_{\Xcal},\alpha_Y,\alpha_{\rm GB})$. This parametrization trivially satisfies all critical conditions.

Then, $\mathcal{L}_2$ and $\mathcal{L}_4^{(2)}$ are given by
\begin{align}
\mathcal{L}_2&=\RA+\frac{1}{2}(\alpha_{T_1}-1)\Ta_{\mu\nu\rho}\Ta^{\mu\nu\rho}-\frac{2}{3\alpha^2}(\alpha_{T_1}-\alpha^2)T_{\mu}T^{\mu}+\frac{3}{2}(\alpha_{T_3}-1)\Tcal_{\mu}\Tcal^{\mu}
\,, \label{GF_L2} \\
\mathcal{L}_4^{(2)}&=\frac{\alpha_{R}}{12}\RA^2+\frac{\alpha_{C}}{4}\CA^2+\frac{\alpha_{\Xcal}}{12}\Xcal^2
+\frac{1}{2}(\alpha_{C}+3\alpha_{Y})Y_{\mu\nu}Y^{\mu\nu}+\alpha_{\rm GB}\RA_{\rm GB}^2
\,, \label{GF_L4}
\end{align}
where
\begin{align}
Y_{\mu\nu}:=\Xs_{\mu\nu}+\frac{1-\alpha}{3\alpha}\FT_{\mu\nu}\,.
\end{align}
The stability conditions of the Minkowski spacetime, i.e.~no ghost and no tachyon conditions on the massive $0^+,0^-,1^+,2^+$ particles, lead to
\begin{align}
\alpha_R,\alpha_C,\alpha_{\Xcal},\alpha_Y>0\,, \quad 0<\alpha^2<\alpha_{T_1}<1\,, \quad \alpha_{T_1}<\alpha_{T_3}
\,. \label{stability_M}
\end{align}
The parameter $\alpha$ can be positive or negative as long as the modulus is less than one but two branches are disconnected since $\alpha\neq 0$ (unless $\alpha_{T_1} = 3\alpha_Y + \alpha_C = 0$).
To analyze cosmological solutions, it is useful to define the parameter 
\begin{align}
\alpha_{T_2}:=\alpha_{T_1}/\alpha^2\,, \label{alphaT2_cond}
\end{align}
where the stability condition \eqref{stability_M} reads
\begin{align}
\alpha_{T_2}>1 \,. \label{alphaT2}
\end{align} 
The masses of the particles around the Minkowski background are
\begin{align}
m^2_{2^+}&=\frac{\alpha_{T_1}}{\alpha_C(1-\alpha_{T_1})}M_*^2 
\,, \\
m^2_{0^+}&=\frac{\alpha_{T_1}}{\alpha_R(\alpha_{T_1}-\alpha^2)}M_*^2=\frac{\alpha_{T_2}}{\alpha_R(\alpha_{T_2}-1)}M_*^2
\,, \\
m^2_{1^+}&=\frac{\alpha_{T_1} \alpha_{T_3}}{\alpha_Y(\alpha_{T_3}-\alpha_{T_1})}M_*^2
\,, \\
m^2_{0^-}&=\frac{\alpha_{T_3}}{\alpha_{\Xcal}}M_*^2\,,
\label{m_0-}
\end{align}
where the suffixes refer to the spin and the parity of each particle species, e.g.~$m_{2^+}$ is the mass of the parity even massive spin-2 particle\footnote{We obtain a qPGT when eliminating $\FT_{\mu\nu}$ from \eqref{GF_L4}.
A case without $\FT_{\mu\nu}$ is $\alpha=1$ where either $2^+$ or $0^+$ is a ghost particle. Another qPGT is the case $\alpha_C+3\alpha_Y=0$ in which either $2^+$ or $1^+$ is a ghost particle. One can tune couplings constant in order that the resultant ghost becomes infinitely heavy and then obtain a ghost-free qPGT with three massive particle species. An example of a ghost-free qPGT is $\alpha=1$ and $\alpha_R=0$ which has massive $2^+,1^+,0^-$ particle species around the Minkowski background. In particular, the cosmological solutions and cosmological perturbations were recently discussed in~\cite{Nikiforova:2016ngy,Nikiforova:2018pdk} under the further assumption $\alpha_{T_3}=\alpha_{T_1}$ where the $1^+$ particle is also infinitely heavy. 
}.

If we introduce a dilaton $\phi$ to restore the local Weyl invariance at all scales, the dimension two parts of the action becomes
\begin{align}
S_2=\int d^4x & \sqrt{-g} \frac{M_{\rm pl}^2}{2}e^{-2\phi}
\nn
&\times \left[ \RA+\frac{1}{2}(\alpha_{T_1}-1)\Ta_{\mu\nu\rho}\Ta^{\mu\nu\rho}-6(\alpha_{T_2}-1) \left(\partial_{\mu}\phi+\frac{1}{3}T_{\mu}\right)^2+\frac{3}{2}(\alpha_{T_3}-1)\Tcal_{\mu}\Tcal^{\mu} \right]
,
\end{align}
where the dilaton is supposed to be transformed as
\begin{align}
\phi \rightarrow \phi +\Omega 
\,, \label{dilaton_trans}
\end{align}
under the Weyl transformation \eqref{Weyl_metric}. By the use of a normalized field 
\begin{align}
\Phi=M_{\rm pl}e^{-\phi}
\,,
\end{align}
the action is written as
\begin{align}
S_2=\int d^4x \sqrt{-g} \left[ -3(\alpha_{T_2}-1) \tilde{\nabla}_{\mu}\Phi \tilde{\nabla}^{\mu}\Phi+\frac{1}{2}\Phi^2 \left( \RA+\frac{1}{2}(\alpha_{T_1}-1)\Ta_{\mu\nu\rho}\Ta^{\mu\nu\rho}+\frac{3}{2}(\alpha_{T_3}-1)\Tcal_{\mu}\Tcal^{\mu} \right) \right]\,,
\end{align}
where the covariant derivative $\tilde{\nabla}_{\mu}$ of $\Phi$ is defined by
\begin{align}
\tilde{\nabla}_{\mu}\Phi=\left(\partial_{\mu}-\frac{1}{3}T_{\mu}\right)\Phi
\,,
\end{align}
which is transformed as
\begin{align}
\tilde{\nabla}_{\mu}\Phi \rightarrow e^{-\Omega} \tilde{\nabla}_{\mu}\Phi 
\,,
\end{align}
under \eqref{Weyl_metric} and \eqref{dilaton_trans}.
We use the same notation as in \eqref{til_nabla} since $\tilde{\nabla}_{\mu}$ is defined in order that a covariant derivative of a tensor $\tilde{\nabla}_{\mu}A_{\cdots}$ is transformed under the local Weyl transformation in the same way as the global one. See, e.g.,~\cite{Blagojevic:2002du} for a discussion on the Weyl gauge theory.
Then, one can see that the parameter $\alpha_{T_2}$ (or $\alpha$ related by \eqref{alphaT2_cond}) has a different physical meaning than $\alpha_{T_1}$ and $\alpha_{T_3}$: $\alpha_{T_2}$ determines the kinetic term of the dilaton. In particular, the limit $\alpha_{T_2}\gg 1$ is thus a weak coupling limit of $\Phi$. In the rest of the present paper, we adopt the unitary gauge
\begin{align}
\Phi=M_{\rm pl} 
\,,
\end{align}
and thus the dilaton does not appear in the Lagrangian explicitly.

In practice, it would be convenient to use an equivalent form of the Lagrangian \eqref{LG} specially for the perturbation analysis.
Since the Gauss-Bonnet term is a boundary term, we can choose $\alpha_{\rm GB}=-\frac{\alpha_C}{4}$ without loss of generality in order to remove $\Ra^2$ term from $\CA^2$:
\begin{align}
\mathcal{L}_4^{(2)}=\frac{2\alpha_R-\alpha_C}{24}\RA^2+\frac{\alpha_C}{2}\RA^{T\mu\nu}\RA^T_{\mu\nu}+\frac{\alpha_{\Xcal}}{12}\Xcal^2
+\frac{1}{2}(\alpha_C+3\alpha_Y )Y_{\mu\nu}Y^{\mu\nu}-\frac{\alpha_C}{2}  {}^{\star}\!\Xcal_{\mu\nu} {}^{\star}\! \Xcal^{\mu\nu}
\,.
\end{align}
Supposing $\alpha_R,\alpha_C,\alpha_{\Xcal},\alpha_Y \neq 0$, we then introduce four dimensionless auxiliary variables $\lambda,\varphi,\Xi_{\mu\nu}$ and $A_{\mu\nu}$ in order to rewrite the action into the equivalent form
\begin{align}
\mathcal{L}_{\rm eq}&=\frac{M_{\rm pl}^2}{2} \biggl[ \lambda \RA +\Xi^{\mu\nu}\GA_{\mu\nu}+\Xcal \varphi+
Y^{\mu\nu} A_{\mu\nu}
\nn 
&\qquad \qquad 
+
\frac{1}{2}(\alpha_{T_1}-1)\Ta_{\mu\nu\rho}\Ta^{\mu\nu\rho}-\frac{2}{3}(\alpha_{T_2}-1)T_{\mu}T^{\mu}+\frac{3}{2}(\alpha_{T_3}-1)\Tcal_{\mu}\Tcal^{\mu} \biggl]
\nn
&+\frac{M_{\rm pl}^2}{M_*^2}\left[ \frac{\alpha_C(1-\alpha)}{3\alpha}\left( Y^{\mu\nu}F_{\mu\nu}-\frac{(1-\alpha)}{6 \alpha} \FT_{\mu\nu}\FT^{\mu\nu} \right)
+\mathcal{L}_4^{(3)}+\mathcal{L}_4^{(4)} \right]
\nn
&- M_{\rm pl}^2 M_*^2 \left[ \frac{3}{4\alpha_R}(\lambda -1)^2 +\frac{1}{8\alpha_C}(\Xi_{\mu\nu}\Xi^{\mu\nu}-\Xi^{\mu}{}_{\mu}\Xi^{\nu}{}_{\nu} ) 
+\frac{3}{4\alpha_{\Xcal} }\varphi^2+\frac{1}{24\alpha_Y}A_{\mu\nu}A^{\mu\nu}\right]\,,
\label{L_eq}
\end{align}
where $\Xi_{\mu\nu}$ and $A_{\mu\nu}$ are symmetric and antisymmetric tensors, respectively.
The original action \eqref{LG} with \eqref{GF_L2} and \eqref{GF_L4} is obtained when integrating out the auxiliary variables  $\lambda,\varphi,\Xi_{\mu\nu}$ and $A_{\mu\nu}$. On the other hand, as shown in Appendix~\ref{appendix}, all components of the torsion can be eliminated by the use of its equation of motion under the assumption $|\RA_{\mu\nu\rho\sigma}| \ll M_*^2, |T_{\mu\nu\rho}| \ll M_*$ and then the Einstein frame action is obtained after transformations of the metric and of the variable $\Xi_{\mu\nu}$. The massive spin-$0^+,0^-,2^+,1^+$ particles are then represented by the scalar, the pseudo-scalar, the symmetric rank two tensor, and the antisymmetric rank two tensor, respectively. 
Although the paper \cite{Aoki:2019snr} only discussed the perturbations around the Minkowski background, the existence of the Einstein frame implies that the theory is free from ghost as far as the curvature and the torsion are sufficiently small in the unit of $M_*$.
The assumption $|\RA_{\mu\nu\rho\sigma}| \ll M_*^2, |T_{\mu\nu\rho}| \ll M_*$, however, does not hold during the inflationary universe. Even so, \eqref{L_eq} is indeed useful to discuss the tensor perturbations as we will see.

We finally specify the nonlinear terms $\mathcal{L}_4^{(3)}$ and $\mathcal{L}_{4}^{(4)}$. They are given by
\begin{align}
\mathcal{L}_4^{(3)}&=c_1 \RA \Tcal_{\mu}\Tcal^{\mu}+c_2 \RA^T_{\mu\nu}\Tcal^{\mu}\Tcal^{\nu}
+c_3\epsilon^{\alpha\beta\rho\sigma}\Ra_{\mu\nu\rho\sigma}\Ta_{\alpha}{}^{\mu\nu}\Tcal_{\beta}
+c_4\Xcal^T_{\mu\nu}\Ta^{\mu\nu\rho}\Tcal_{\rho}
+c_5\epsilon^{\mu\alpha\beta\gamma}\RA^T_{\mu\nu}\Ta_{\alpha\beta}{}^{\nu}\Tcal_{\gamma}
\nn
&+c_6 \epsilon^{\alpha\beta\gamma\mu}\Xs_{\mu\nu}\Ta_{\alpha\beta}{}^{\nu}\Tcal_{\gamma}
+c_7 \epsilon^{\alpha\beta\gamma\mu}\FT_{\mu\nu}\Ta_{\alpha\beta}{}^{\nu}\Tcal_{\gamma}
+O(\Ta^2)
\nn
&=c_1 \RA \Tcal_{\mu}\Tcal^{\mu}+c_2 \RA^T_{\mu\nu}\Tcal^{\mu}\Tcal^{\nu}
+c_3\Ra_{\mu\nu\rho\sigma}\Ta^{\alpha\mu\nu}\Tc^{\rho\sigma}{}_{\alpha}
-\frac{1}{2}c_4\Rb_{\mu\nu\rho\sigma}\Ta^{\alpha\mu\nu}\Tc^{\rho\sigma}{}_{\alpha}
\nn
&+\frac{1}{2}c_5\Rd_{\mu\nu\rho\sigma}\Ta^{\alpha\mu\nu}\Tc^{\rho\sigma}{}_{\alpha}
+\frac{1}{2}c_6\Ree_{\mu\nu\rho\sigma}\Ta^{\alpha\mu\nu}\Tc^{\rho\sigma}{}_{\alpha}
+c_7 \FT_{\mu\nu}\Ta^{\mu\alpha\beta}\Tc^{\nu}{}_{\alpha\beta}
+O(\Ta^2)
\,, \label{L_4^3} \\
\mathcal{L}_4^{(4)}&=d_1 (\Tcal_{\mu}\Tcal^{\mu})^2+O(\Ta^2)
\,, \label{L_4^4}
\end{align}
up to linear order in $\Ta_{\mu\nu\rho}$. In principle, there would be no reason to exclude terms nonlinear in $\Ta_{\mu\nu\rho}$ such as $\RA \Ta_{\mu\nu\rho}\Ta^{\mu\nu\rho}$ and $ \Ta^{\mu\nu\rho}\Ta_{\mu\nu\rho}\Tcal^{\alpha}\Tcal_{\alpha}$. However, we do not consider such terms in the present paper since they do not affect the background dynamics of the universe for the same reason to ignore $\nablaA_{\mu}\Ta^{\nu}{}_{\rho\sigma}$. The linear terms are added to just show that they contribute to the perturbations and can be constrained. For instance, in Sec.~\ref{sec_tensor}, we will find that the ghost-free condition of the tensor perturbations in the high momentum limit leads to $c_3=c_4=0$.


\section{Einstein frame in single field limit}
\label{sec_heavy}
As a first step to study the inflationary universe, we may consider a parameter space where only the inflaton and the graviton are light fields while masses of other fields are much larger than the Hubble scale. Here, we identify the spin-$0^+$ particle with the inflaton field and consider the limit $\alpha_C,\alpha_Y,\alpha_{\Xcal}\rightarrow 0$, i.e. heavy mass limit of the $2^+,1^+,0^-$ particle species. To consider this limit, we first use the original Lagrangian \eqref{LG}. The equations of motion of $\Ta^{\mu}{}_{\nu\rho}$ and $\Tcal_{\mu}$ generally admit a solution
\begin{align} 
\Ta^{\mu}{}_{\nu\rho}= 0\,, \quad \Tcal_{\mu} = 0
\,, \label{T1_Tc=0}
\end{align}
under $\alpha_C,\alpha_Y,\alpha_{\Xcal}\rightarrow 0$.
In the infinite mass limit, fluctuations of $\Ta^{\mu}{}_{\nu\rho}, \Tcal_{\mu}$ can be ignored as far as the solution \eqref{T1_Tc=0} is stable; we thus consider the action
\begin{align}
\mathcal{L}_{\rm light}
&=\frac{M_{\rm pl}^2}{2}\left[\RA-\frac{2}{3}(\alpha_{T_2}-1)T_{\mu}T^{\mu}+\frac{\alpha_R}{6M_*^2}\RA^2 \right]_{\Ta,\Tcal=0}\,. \label{L_light}
\end{align}

We then rewrite the action \eqref{L_light} as the equivalent form
\begin{align}
\sqrt{-g}\mathcal{L}_{\rm eq}=\sqrt{-g}\left[\frac{M_{\rm pl}^2}{2}\lambda \RA|_{\Ta,\Tcal=0}-\frac{M_{\rm pl}^2}{3}(\alpha_{T_2}-1)T_{\mu}T^{\mu}-\frac{3M_{\rm pl}^2M_*^2}{4\alpha_R}(\lambda-1)^2\right]\,, \label{L_light_with_lambda}
\end{align}
where $\lambda$ is an auxiliary variable. In the similar way as what is done in the Starobinsky inflation, we can take a Weyl transformation of \eqref{L_light_with_lambda} to obtain the Einstein frame action. 
In the present case, the computation is simpler to consider the following Weyl transformation
\begin{align}
g_{\mu\nu}=\lambda^{-1}g^E_{\mu\nu}\,,\quad T_{\mu}=T_{\mu}^E+\frac{3}{2}\partial_{\mu} \ln \lambda
\,, \label{metric_trans}
\end{align}
with $\lambda>0$.
The Einstein frame action is
\begin{align}
\sqrt{-g}\mathcal{L}_{\rm eq}
&=\sqrt{-g^E}\left[ \frac{M_{\rm pl}^2}{2}\RA_E|_{\Ta,\Tcal=0}-\frac{M_{\rm pl}^2}{3\lambda}(\alpha_{T_2}-1)\left(T_{\mu}^E+\frac{3}{2}\partial_{\mu} \ln \lambda \right)^2
-\frac{3M_{\rm pl}^2 M_*^2}{4\alpha_R}(1-\lambda^{-1})^2 \right]\,,
\end{align}
where
\begin{align}
\RA_E|_{\Ta,\Tcal=0}&=R(g^E)-\frac{2}{3}T_{\mu}^E T^{E\mu} -2\nabla^E_{\mu}T^{E\mu}\,.
\end{align}
The Einstein frame Ricci scalar $R(g^E)$ and the covariant derivative $\nabla^E_{\mu}$ are defined by the Levi-Civita connection with respect to $g^E_{\mu\nu}$. We can further eliminate $T^E_{\mu}$ by using its equation of motion of which solution is
\begin{align}
T^E_{\mu}=-\frac{3(\alpha_{T_2}-1)}{2(\alpha_{T_2}-1+\lambda)}\partial_{\mu} \ln \lambda
\,.
\end{align}
The Einstein frame action is then
\begin{align}
\sqrt{-g^E}\mathcal{L}_E=\sqrt{-g^E}\left[ \frac{M_{\rm pl}^2}{2}R(g^E)-\frac{3M_{\rm pl}^2(\alpha_{T_2}-1)}{4(\alpha_{T_2}-1+ \lambda)} (\partial_{\mu} \ln \lambda)^2-\frac{3M_{\rm pl}^2 M_*^2}{4\alpha_R} (1-\lambda^{-1})^2 \right]\,.
\end{align}

One can immediately see that the Starobinsky's inflationary model is recovered in the limit $\alpha_{T_2}\rightarrow \infty$:
\begin{align}
\lim_{\alpha_{T_2}\rightarrow 0} \mathcal{L}_E
&=\frac{M_{\rm pl}^2}{2}R(g^E)-\frac{3M_{\rm pl}^2}{4} (\partial_{\mu} \ln \lambda)^2-\frac{3M_{\rm pl}^2 M_*^2}{4\alpha_R} (1-\lambda^{-1})^2
\nn
&=\frac{M_{\rm pl}^2}{2}R(g^E)-\frac{1}{2}(\partial_{\mu} \chi_0)^2-\frac{3M_{\rm pl}^2 M_*^2}{4\alpha_R}\left(1-e^{-\sqrt{\frac{2}{3}} \frac{\chi_0}{M_{\rm pl}} } \right)^2
\,,
\end{align}
where the canonically normalized field $\chi_0$ is defined via
\begin{align}
\lambda = e^{\sqrt{\frac{2}{3}} \frac{\chi_0}{M_{\rm pl}} }
\,. \label{canonical_Sta}
\end{align}
The restoration of the Starobinsky model is easily understood by the fact that the equation of motion of $T_{\mu}$ yields $T_{\mu}\rightarrow 0$ in the limit $\alpha_{T_2}\rightarrow \infty$ which is nothing but the limit to the torsionless geometry, namely the Riemannian geometry. The Starobinsky model is obtained by taking $\alpha_C,\alpha_{\Xcal},\alpha_Y \rightarrow 0$, namely the infinitely heavy mass limit of $2^+,1^+,0^-$ and the weak coupling limit of the dilaton, $\alpha_{T_2}\rightarrow \infty$.

\begin{figure}[tbp]
\centering
\includegraphics[width=7cm,angle=0,clip]{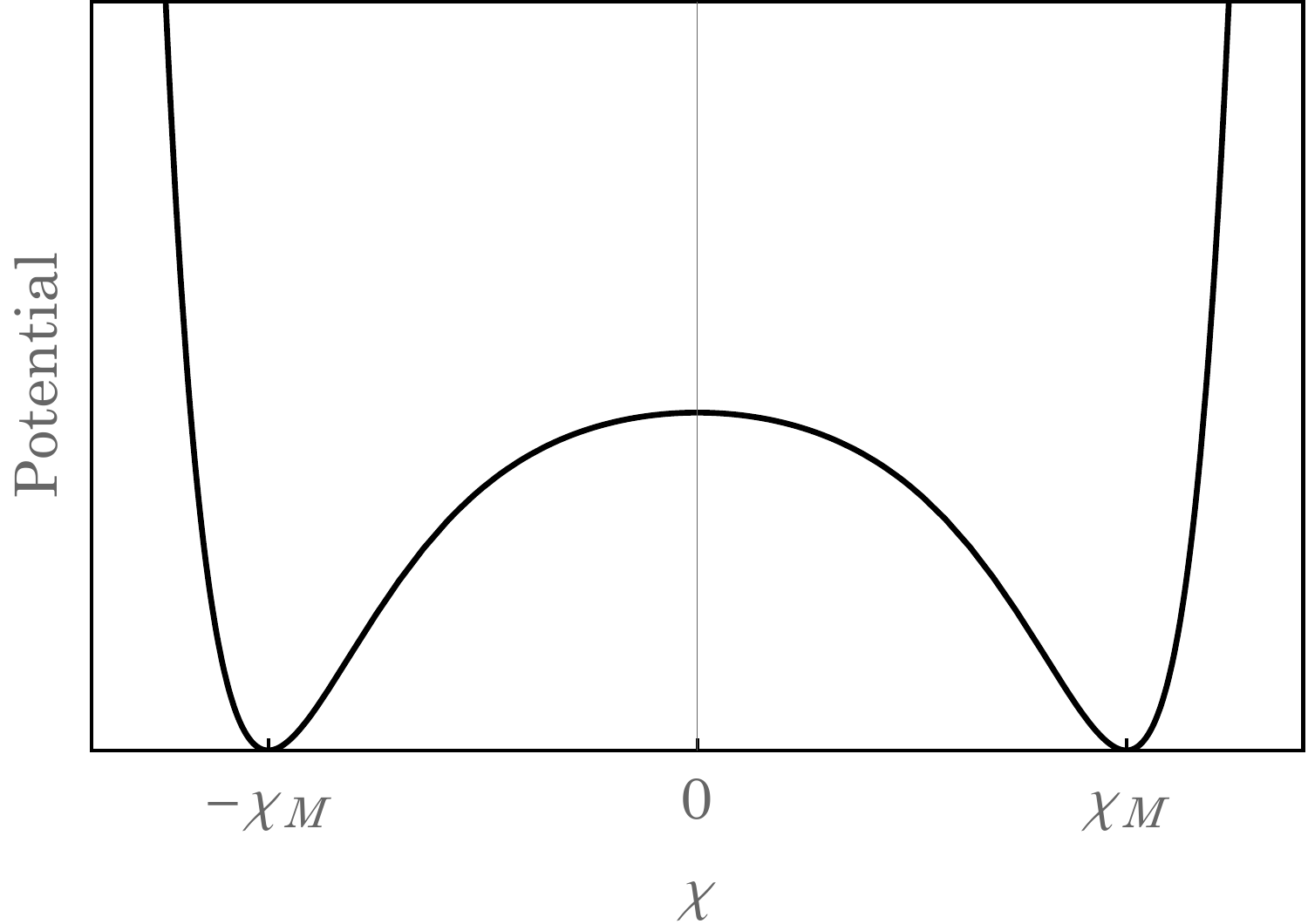}
\qquad
\includegraphics[width=7cm,angle=0,clip]{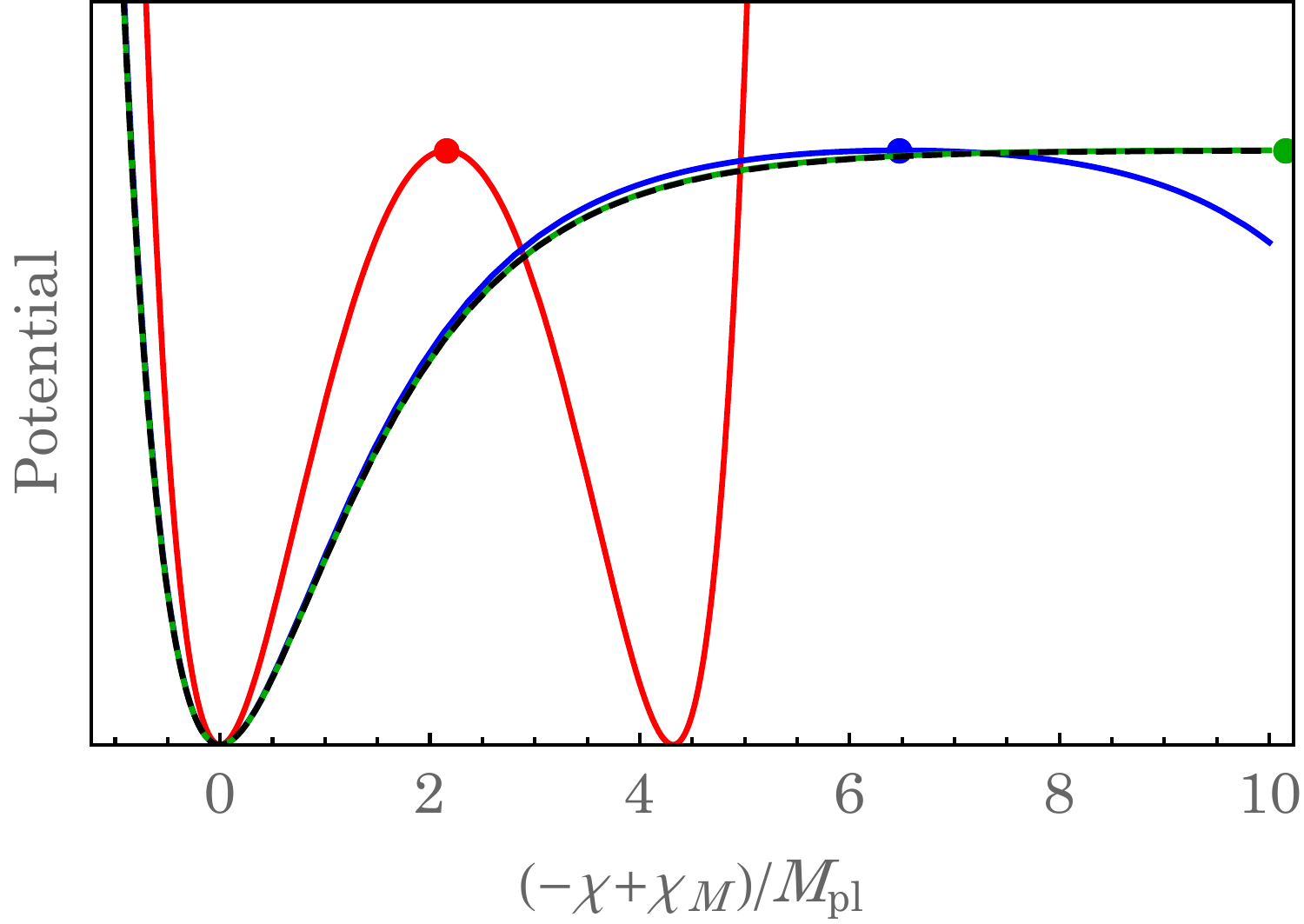}
\caption{The form of the potential. In the right figure, we set $\alpha_{T_2}=2$ (red), $\alpha_{T_2}=50$ (blue), and $\alpha_{T_2}=1000$ (green), where each dot is the local maximum of the potential. The black dashed curve in the right figure is the form of the Starobinsky potential.
}
\label{fig_V0}
\end{figure}

For a generic value of $\alpha_{T_2}$, we introduce a canonically normalized field $\chi$ via 
\begin{align}
\lambda=(\alpha_{T_2}-1) {\rm csch^2}\left[ \frac{\chi}{\sqrt{6} M_{\rm pl}} \right]\,.
 \label{canonical}
\end{align}
The Einstein frame action with the canonically normalized field $\chi$ is given by
\begin{align}
\mathcal{L}_E=\frac{M_{\rm pl}^2}{2}R(g^E)-\frac{1}{2}(\partial_{\mu} \chi)^2-V(\chi)\,, \label{Einstein_single}
\end{align}
with the potential
\begin{align}
V(\chi)=\frac{3M_{\rm pl}^2 M_*^2}{16\alpha_R(\alpha_{T_2}-1)^2} \left( (2\alpha_{T_2}-1)- \cosh\left[\sqrt{\frac{2}{3}} \frac{\chi}{M_{\rm pl}} \right] \right)^2\,, \label{potential_V}
\end{align}
which is an even function of $\chi$. The form of the potential is shown in the left panel of Fig.~\ref{fig_V0}, and the overall shape including the positive and negative ranges of $\chi$ is similar to that of the hilltop inflation~\cite{Boubekeur:2005zm} rather than the Starobinsky inflation. However, we should only discuss the range $\chi>0$ since $\chi=0$ is a singular point of the transformation \eqref{metric_trans}. When considering the cosmology, the point $\chi=0$ is the Big Bang singularity where the scale factor of the original metric $g_{\mu\nu}$ vanishes. It would be nonetheless interesting that the point $\chi=0$ is not a singularity in the Einstein frame action. The point $\chi=0$ is just a local maximum of the potential where the unstable de Sitter spacetime is an exact solution.

The Minkowski spacetime is realized at $\chi=\chi_M$ ($\lambda=1$), where 
\begin{align}
\chi_M=\sqrt{\frac{3}{2}}M_{\rm pl} {\rm arccosh} (2\alpha_{T_2}-1 )\,.
\end{align}
Two variables $\chi_0,\chi$ defined by \eqref{canonical_Sta} and \eqref{canonical} are related via
\begin{align}
\chi_0 =\lim_{\alpha_{T_2} \rightarrow \infty}(-\chi+\chi_M)
\,.
\end{align} 
We notice that $\chi_M \rightarrow \infty$ as $\alpha_{T_2} \rightarrow \infty$. Therefore, under the limit $\alpha_{T_2}\rightarrow \infty $, the singularity is at $\chi=0$ and $\chi_0=\infty$ while the Minkowski point is at $\chi=\infty$ and $\chi_0=0$, respectively.
From the right panel of Fig.~\ref{fig_V0}, one can see that the potential indeed approaches the Starobinsky potential as $\alpha_{T_2}\rightarrow \infty$ and that the singular point (the local maximum of the potential) goes to infinity as $\alpha_{T_2}\rightarrow \infty$.

\begin{figure}[tbp]
\centering
\includegraphics[width=7cm,angle=0,clip]{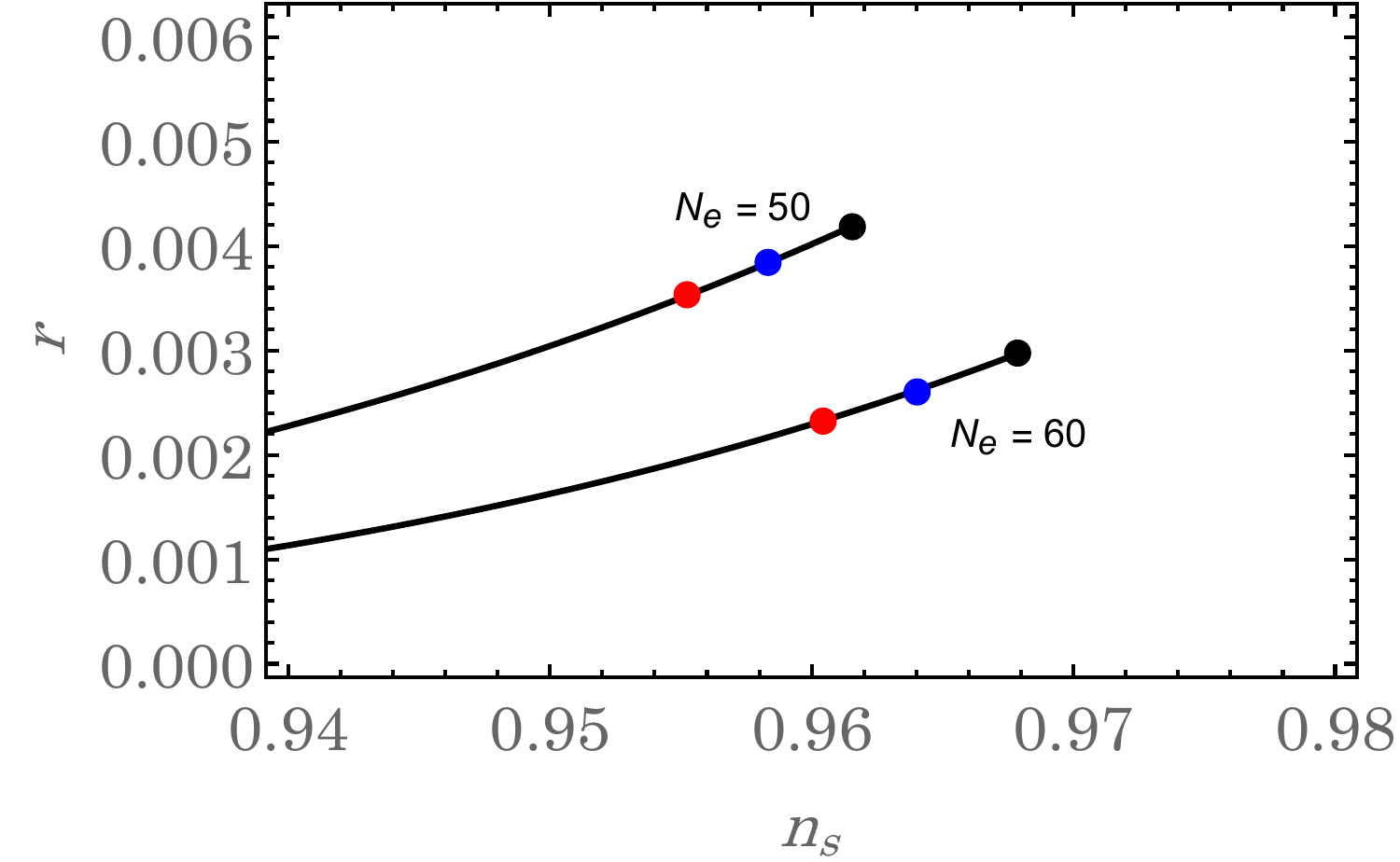}
\caption{The values of $n_s$ and $r$, where the dots corresponds to the results of $\alpha_{T_2}=50$ (red), $\alpha_{T_2}=70$ (blue) and the Starobinsky inflation (black).
}
\label{fig_ns_r}
\end{figure}

Since the Einstein frame action is given by GR with a canonical scalar field with a potential, we may use the standard slow-roll approximation. The slow-roll parameters are
\begin{align}
\epsilon_V&:=\frac{M_{\rm pl}^2}{2}\left(\frac{dV/d\chi}{V}\right)^2
=\frac{4(y+1)(y-1)}{3[(y+1)-2\alpha_{T_2}]^2}
\,, 
\label{ep_analytic} \\
\eta_V&:=M_{\rm pl}^2 \frac{d^2V/d\chi^2}{V}
=\frac{4[2y^2-(2\alpha_{T_2}-1)y-1]}{3[(y+1)-2\alpha_{T_2}]^2}
\,, \label{eta_analytic}
\end{align}
where
\begin{align}
y:=\cosh\left[\sqrt{\frac{2}{3}} \frac{\chi}{M_{\rm pl}} \right]\,.
\label{def_y}
\end{align}
The e-folds is
\begin{align}
N_e&\simeq \frac{3}{2}\int_{\chi_e}^{\chi}d\chi' \frac{V_0}{dV_0/d\chi'}
\nn
&=\left[ \frac{3}{4}\left\{-(\alpha_{T_2}-1)\ln(y-1)+\alpha_{T_2} \ln(y+1) \right\} \right]^{y}_{y_e}\,,
\label{Ne_analytic}
\end{align}
where $\chi_e$ and $y_e$ are the values of $\chi$ and $y$, respectively, at the end of inflation, i.e. a root of $\epsilon_V=1$. 
The asymptotic value of $y_e$ as $\alpha_{T_2} \rightarrow \infty$ is $y_e \rightarrow (4\sqrt{3}-6)\alpha_{T_2} \simeq 0.93 \alpha_{T_2}$. The contribution from the lower limit of the integral logarithmically increases as $\alpha_{T_2}$ increases. Therefore, as long as $\frac{3}{4} \ln \alpha_{T_2} \ll N_e$, we may ignore the contribution from the lower limit and obtain
\begin{align}
N_e &\simeq -\frac{3(\alpha_{T_2}-1)}{4}\ln(y-1)+\frac{3\alpha_{T_2}}{4}\ln(y+1) 
\nn
&=\frac{3\alpha_{T_2}}{4} \ln \left( \frac{y+1}{y-1} \right)+\frac{3}{4}\ln(y-1) \,.
\end{align}
This expression implies that the parameter $\alpha_{T_2}$ controls how long $\chi$ can stay on the slope of the potential where we shall call the region $y\simeq 1$ the top of the potential and $\alpha_{T_2} \gtrsim y-1>O(1)$ the slope of the potential, respectively. 
First, let us consider the case with $y\simeq 1$, i.e. the case where the observed CMB scale corresponds to the region near the top of the potential. We obtain
\begin{align}
y \simeq 1+\exp \left[- \frac{4N_e}{3(\alpha_{T_2}-1)} \right]\,,
\label{y_top}
\end{align}
and then
\begin{align}
\epsilon_V &\simeq \frac{2}{3(\alpha_{T_2}-1)^2}\exp\left[ -\frac{4N_e}{3(\alpha_{T_2}-1)} \right]
\,, \\
\eta_V &\simeq -\frac{2}{3(\alpha_{T_2}-1)}\,,
\end{align}
where the approximation is valid when $\frac{4N_e}{3(\alpha_{T_2}-1)} \gg 1$.
This scenario must be excluded by the current observations since $\epsilon_V \ll 1, |\eta_V| =O(1)$ and then the spectral index $n_s$ largely deviates from 1. The situation is similar to the hilltop inflation. Hence, we consider the case $\alpha_{T_2} \gtrsim N_e (\simeq 50) \gg 1$ where the observed scale is on the slope of the potential, $y-1> O(1)$. In this case, we have
\begin{align}
y\simeq \frac{e^{\frac{4N_e}{3\alpha_{T_2}} } +1}{e^{\frac{4N_e}{3\alpha_{T_2}} } -1 }\,,
\label{y_slope}
\end{align}
and then we obtain
\begin{align}
\epsilon_V &\simeq \frac{4}{3\alpha_{T_2}^2}e^{\frac{4N_e}{3\alpha_{T_2}} } \left(e^{\frac{4N_e}{3\alpha_{T_2}} }-1 \right)^{-2}\,, \label{epsilon_V} \\
\eta_V &\simeq \frac{2}{3\alpha_{T_2}}\left( 1- e^{\frac{8N_e}{3\alpha_{T_2}} } \right) \left(e^{\frac{4N_e}{3\alpha_{T_2}} }-1 \right)^{-2}\,. \label{eta_V}
\end{align}
In particular, when $\alpha_{T_2} \gg N_e$, the slow roll parameters can be approximated as
\begin{align}
\epsilon_V &\simeq \frac{3}{4N_e^2}
\,, \\
\eta_V &\simeq -\frac{1}{N_e}\,,
\end{align}
which reproduce the predictions of the Starobinsky inflation. Fig.~\ref{fig_ns_r} shows numerical values of the spectral index $n_s$ and the tensor-to-scalar ratio $r$ computed by \eqref{ep_analytic}, \eqref{eta_analytic}, and \eqref{Ne_analytic}. Although the precise constraint on $\alpha_{T_2}$ depends on the e-folds, the result yields the lower bound, $\alpha_{T_2} \gtrsim 50$; for instance, $\alpha_{T_2}=50,N_e=55$ gives $n_s=0.958$ and $r=0.0028$ which is almost the lower bound of $n_s$ at $2\sigma$ level~\cite{Akrami:2018odb}.

An interesting observation would be that the tensor-to-scalar ratio is almost unchanged in the viable range of $n_s$. The tensor-to-scalar ratio is of order $10^{-3}$ not only for the Starobinsky model but also for the $R^2$ model based on the Riemann-Cartan geometry with finite $\alpha_{T_2}$. Therefore, if the future observations such as LiteBIRD do not detect the primordial gravitational waves, both of the $R^2$ models are excluded. On the other hand, if the tensor-to-scalar ratio as well as $n_s$ is found to be consistent with the predictions of the $R^2$ models, more precise observations may be used to explore the spacetime geometry during the inflationary regime.

How can we distinguish between the Riemannian geometry and the Riemann-Cartan geometry if the inflation model is the $R^2$ model? The crucial difference between them would be the number of the non-ghost massive particle species: there is only the massive spin-$0^+$ particle in the Riemannian case while there can be four massive particle species in the Riemann-Cartan case. Signatures of these massive particle species must be a smoking gun to distinguish these models which are discussed in the following sections.

\section{Background dynamics in general case}
\label{sec_BG}
In  this section, we study dynamics of the flat FLRW universe
\begin{align}
ds^2=-N^2 d\tau^2+a^2d\bm{x}^2\,, \quad T_{\mu}dx^{\mu}=N T_0 d\tau \,,\quad 
\Tcal_{\mu}dx^{\mu}=N \Tcal_0 d\tau \,,
\end{align}
in the general parameter space of \eqref{LG}, where all variables are functions of the time $\tau$.

\subsection{Non-zero operators for minisuperspace action}
In the flat FLRW ansatz, we have
\begin{align}
\Ta^{\mu}{}_{\nu\rho}=0\,,\quad \FT_{\mu\nu}=0\,, \quad \tilde{\nabla}_{\mu}\Ta^{\nu}{}_{\rho\sigma}=0 \,, \quad 
\Ra_{\mu\nu\rho\sigma}=0
\,.
\end{align}
These tensors do not contribute the background dynamics of the universe. Only non-zero components of the Riemann-Cartan curvature are given by
\begin{align}
\RA^{0}{}_{0}&=\frac{3H'}{N}+3H^2+\frac{T_0'}{N}+T_0H
\,, \\
\RA^{i}{}_{j}&=\left[\frac{H'}{N}+3H^2+\frac{T_0'}{3N}+\frac{5}{3}HT_0+\frac{2}{9}T_0^2-\frac{1}{2}\Tcal_0^2\right] \delta^i_j
\,, \\
\Xcal^0{}_{0}&=-(3H+T_0)\Tcal_0 
\,, \\
\Xcal^i{}_j&=-\left[\frac{\Tcal_0'}{N}+\left(2H+\frac{1}{3}T_0\right)\Tcal_0\right]\delta^i_j
\,,
\end{align}
where $H:=\frac{a'}{aN}$ and the prime is the derivative with respect to $\tau$. From the parity invariance as well as the local Weyl invariance, the most general non-vanishing operator at the scaling dimension four is thus a linear combination of
\begin{align}
\RA^2,~ \RA^T_{\mu\nu}\RA^{T\mu\nu},~ \Xcal^2,~ \Xcal^T_{\mu\nu}\Xcal^{T\mu\nu},~ \RA \Tcal_{\mu}\Tcal^{\mu}\,,~ \RA^T_{\mu\nu}\Tcal^{\mu}\Tcal^{\nu},~ (\Tcal_{\mu}\Tcal^{\mu})^2\,, \label{L4_FLRW}
\end{align}
in the FLRW ansatz. The non-vanishing part of the Lagrangian \eqref{LG} with \eqref{GF_para} and \eqref{alphaT2_cond} under the FLRW ansatz is
\begin{align}
\mathcal{L}_{\rm FLRW}&=\frac{M_{\rm pl}^2}{2}\left[\RA-\frac{2}{3}\left(\alpha_{T_2}-1 \right)T_{\mu}T^{\mu}+ \frac{3}{2}(\alpha_{T_3}-1) \Tcal_{\mu}\Tcal^{\mu} \right]
\nn
&+\frac{M_{\rm pl}^2}{M_*^2}\left[ \frac{\alpha_R}{12}\RA^2+\frac{\alpha_C}{4}\CA^2+\frac{\alpha_{\Xcal}}{12}\Xcal^2 
+c_1 \RA \Tcal_{\mu}\Tcal^{\mu}+c_2 \RA^T_{\mu\nu}\Tcal^{\mu}\Tcal^{\nu}+d_1 (\Tcal_{\mu}\Tcal^{\mu})^2 \right]\,.
\label{L_FLRW_C}
\end{align}
The dynamics of the background universe is determined by $\alpha_{T_2}, \alpha_{T_3},\alpha_R,\alpha_C,\alpha_{\Xcal},c_1,c_2,d_1$.
However, one of $(\alpha_C,c_1,c_2)$ is redundant for discussing the FLRW universe since the combination
\begin{align}
\sqrt{-g}\left(\CA^2+2\RA_{\mu\nu}\Tcal^{\mu}\Tcal^{\nu}\right)&
=\sqrt{-g}\left(2\Xcal^T_{\mu\nu}\Xcal^{T\mu\nu}-\frac{1}{6}\Xcal^2+2\RA_{\mu\nu}\Tcal^{\mu}\Tcal^{\nu}\right)
\nn
&=-\frac{d}{d\tau}\left[ 2N a^3 \Tcal_0^2 (3H+T_0) \right]\,,
\label{bound_RTT}
\end{align}
is a total divergence term for the FLRW ansatz although this is not the case in general. If we remove $\CA^2$ from the minisuperspace action by using \eqref{bound_RTT}, we obtain
\begin{align}
\mathcal{L}_{\rm FLRW}&=\frac{M_{\rm pl}^2}{2}\left[\RA-\frac{2}{3}\left(\alpha_{T_2}-1 \right)T_{\mu}T^{\mu}+ \frac{3}{2}(\alpha_{T_3}-1) \Tcal_{\mu}\Tcal^{\mu} \right]
\nn
&+\frac{M_{\rm pl}^2}{M_*^2}\left[ \frac{\alpha_R}{12}\RA^2+\frac{\alpha_{\Xcal}}{12}\Xcal^2 
+\left(c_1-\frac{\alpha_C}{8} \right) \RA \Tcal_{\mu}\Tcal^{\mu}+\left(c_2-\frac{\alpha_C}{2} \right) \RA^T_{\mu\nu}\Tcal^{\mu}\Tcal^{\nu}+d_1 (\Tcal_{\mu}\Tcal^{\mu})^2 \right]\,.
 \label{L_FLRW}
\end{align}

It would be worth emphasizing that the same conclusion is obtained even if we do not use the ghost-free conditions \eqref{GF_para}. The most general minisuperspace Lagrangian with the parity invariance and the asymptotic Weyl invariance is computed by
\begin{align}
\mathcal{L}^{\rm general}_{\rm FLRW}
&=\frac{M_{\rm pl}^2}{2}\left(\RA+a_2T_{\mu}T^{\mu}+ a_3 \Tcal_{\mu}\Tcal^{\mu} \right)
\nn
& 
+\frac{M_{\rm pl}^2}{M_*^2}\biggl(
-2b_2 \Xcal^T_{\mu\nu}\Xcal^{T\mu\nu}-\frac{b_3}{6}\Xcal^2+2b_4\RA^T_{\mu\nu}\RA^{T\mu\nu}+\frac{b_6}{6}\RA^2
\nn
&\qquad \qquad
+c_1 \RA \Tcal_{\mu}\Tcal^{\mu}+c_2 \RA^T_{\mu\nu}\Tcal^{\mu}\Tcal^{\nu}+d_1 (\Tcal_{\mu}\Tcal^{\mu})^2
\biggl)
\,, \label{Jordan_frame}
\end{align}
by the use of the FLRW ansatz. One can then add the Gauss-Bonnet term $\RA^2_{\rm GB}$ and the boundary term \eqref{bound_RTT} to eliminate two of the terms in the second line of \eqref{Jordan_frame}. As a result, the general minisuperspace action \eqref{Jordan_frame} is given by only eight independent terms. Eliminating $\RA^T_{\mu\nu}\RA^{T\mu\nu}$ and $\Xcal^T_{\mu\nu}\Xcal^{T\mu\nu}$ we obtain the same form as \eqref{L_FLRW},
\begin{align}
\mathcal{L}^{\rm general}_{\rm FLRW}
&=\frac{M_{\rm pl}^2}{2}\left(\RA+a_2T_{\mu}T^{\mu}+ a_3 \Tcal_{\mu}\Tcal^{\mu} \right)
\nn
&+\frac{M_{\rm pl}^2}{M_*^2}\biggl[ \frac{1}{6}(b_4+b_6) \RA^2 -\frac{1}{6}(b_2+b_3) \Xcal^2 +\left(c_1+\frac{b_2}{2}-\frac{b_4}{2}\right) \RA \Tcal_{\mu}\Tcal^{\mu} 
\nn
& \qquad \qquad
+ (c_2+2b_2-2b_4) \RA^T_{\mu\nu}\Tcal^{\mu}\Tcal^{\nu} +d_1 (\Tcal_{\mu}\Tcal^{\mu})^2 \biggl]\,.
\end{align}
Regardless of using the ghost-free parametrization \eqref{GF_para}, there are only eight independent terms in the FLRW ansatz.

\subsection{Quasi-Einstein frame action}
As discussed, the existence of $0^+, 0^-$ particles (and $2^+,1^+$ particles) are made clear by introducing auxiliary variables and using the equivalent form \eqref{L_eq}.
Since only the spin-0 sectors are dynamical in the minisuperspace action, we only introduce two auxiliary scalars $\lambda,\varphi$ (i.e.~integrating out $\Xi_{\mu\nu}$ and $A_{\mu\nu}$ from \eqref{L_eq}) and rewrite the action as
\begin{align}
\mathcal{L}_{\rm FLRW}&=\frac{M_{\rm pl}^2}{2}\left[ \lambda \RA+\varphi \Xcal-\frac{2}{3}\left(\alpha_{T_2}-1 \right)T_{\mu}T^{\mu}+ \frac{3}{2}(\alpha_{T_3}-1) \Tcal_{\mu}\Tcal^{\mu} \right]
\nn
&-\frac{3M_{\rm pl}^2M_*^2}{4\alpha_R}(\lambda-1)^2-\frac{3M_{\rm pl}^2 M_*^2}{2\alpha_{\Xcal}} \varphi^2
\nn
&+\frac{M_{\rm pl}^2}{M_*^2}\left[ 
 \frac{\alpha_C}{4}\CA^2+ c_1 \RA \Tcal_{\mu}\Tcal^{\mu}+c_2 \RA^T_{\mu\nu}\Tcal^{\mu}\Tcal^{\nu}+d_1 (\Tcal_{\mu}\Tcal^{\mu})^2 \right]\,. \label{mini_two_scalar}
\end{align}
One can perform integration by parts in order that the action does not contain time derivatives of $T_0$. Since the action depends on $T_0$ at most quadratically, the constraint equation of $T_0$ is easily solved. After integrating out $T_0$ and taking the redefinition of $\lambda$ as
\begin{align}
\lambda \rightarrow \lambda+\frac{1}{2}(4c_1+c_2-\alpha_C)\Tcal_0^2/M_*^2\,,
\end{align}
the variable $\Tcal_0$ becomes a non-dynamical variable. Therefore, $\Tcal_0$ can, in principle, be integrated out and then the minisuperspace action with the parity even scalar $\lambda$ and the parity odd scalar $\varphi$ would be obtained. In practice, however, the constraint equation of $\Tcal_0$ cannot be easily solved. One could not find the Einstein frame of the minisuperspace action with two scalar fields in generic parameter space.

Therefore, we integrate out $\varphi$ from \eqref{mini_two_scalar} and write the minisuperspace action as
\begin{align}
\mathcal{L}_{\rm FLRW}&=\frac{M_{\rm pl}^2}{2}\left[ \lambda \RA-\frac{2}{3}\left(\alpha_{T_2}-1 \right)T_{\mu}T^{\mu}+ \frac{3}{2}(\alpha_{T_3}-1) \Tcal_{\mu}\Tcal^{\mu} \right]-\frac{3M_{\rm pl}^2M_*^2}{4\alpha_R}(\lambda-1)^2
\nn
&+\frac{M_{\rm pl}^2}{M_*^2}\left[ \frac{\alpha_C}{4}\CA^2+\frac{\alpha_{\Xcal}}{12}\Xcal^2 
+c_1 \RA \Tcal_{\mu}\Tcal^{\mu}+c_2 \RA^T_{\mu\nu}\Tcal^{\mu}\Tcal^{\nu}+d_1 (\Tcal_{\mu}\Tcal^{\mu})^2 \right]\,.
\end{align}
After integrating out $T_0$, we obtain the minisuperspace action in terms of $(a,N,\lambda,\Tcal_0)$. We then take the field redefinitions, $(a,N,\lambda) \rightarrow (a_E,N_E,\tilde{\lambda} )$, via the relations
\begin{align}
a&=A^{-1/2}  a_E\,, \label{change} \\
N&=A^{-1/2}N_E \,, \\
\lambda&=A+\frac{1}{4}(8c_1+2c_2+3\alpha_{\Xcal}-2\alpha_C)\frac{\Tcal_0^2}{M_*^2}\,,
\end{align}
where
\begin{align}
A=\tilde{\lambda}\left(1+\frac{\alpha_{\Xcal}\Tcal_0^2 }{4 M_*^2(\alpha_{T_2}-1)}\right)\,. \label{conformal_A}
\end{align}
The minisuperspace action with the variables $\tilde{\chi}^a=(\tilde{\lambda},\Tcal_0)$ is given by
\begin{align}
S_{\rm FLRW}=\int dt_E d^3\bm{x} a_E^3 \left[ \frac{M_{\rm pl}^2}{2}R_{\rm FLRW}(N_E,a_E)+\frac{1}{2}G_{ab}\frac{d \tilde{\chi}^a}{d t_E}\frac{d \tilde{\chi}^b}{d t_E}-V_{FLRW}(\tilde{\chi}_a) + \mathcal{L}_{\rm non} \right]
\,, \label{Einstein_frame}
\end{align}
where the time $t_E$ is defined by $dt_E=N_E d\tau$ and $R_{\rm FLRW}$ is the Riemannian Ricci scalar,
\begin{align}
R_{\rm FLRW}(N_E,a_E)=12H_E^2+6\frac{dH_E}{dt_E},
\end{align}
with $H_E=\frac{d \ln a}{dt_E}$.
The field space metric and the potential are given by
\begin{align}
G_{ab}={\rm diag}\left[ \frac{3 M_{\rm pl}^2(\alpha_{T_2}-1)}{2\tilde{\lambda}^2(\alpha_{T_2}-1+\tilde{\lambda})}, \frac{24\alpha_{\Xcal} M_{\rm pl}^2 M_*^2(\alpha_{T_2}-1)^2}{\tilde{\lambda} (\alpha_{\Xcal} \Tcal_0^2+4(\alpha_{T_2}-1) M_*^2)^2} \right]\,,
\label{G_metric}
\end{align}
and
\begin{align}
V_{\rm FLRW}=\frac{3M_{\rm pl}^2 M_*^2}{4\alpha_R} (1-\tilde{\lambda}^{-1})^2+\frac{3(\alpha_{T_2}-1)M_{\rm pl}^2 M_*^2}{2\alpha_{\Xcal} \tilde{\lambda}^2}f(\Tcal_0)\left[\alpha_{T_3}+g_1(\tilde{\lambda}-1)+g_2(\alpha_{T_2}-1) f(\Tcal_0) \right]\,,
\end{align}
with the function
\begin{align}
f(\Tcal_0):=\frac{2\alpha_{\Xcal} \Tcal_0^2}{\alpha_{\Xcal} \Tcal_0^2+4(\alpha_{T_2}-1)M_*^2 }\,. \label{def_f}
\end{align}
The last term $\mathcal{L}_{\rm non}$ represents the remaining non-minimal coupling
\begin{align}
\mathcal{L}_{\rm non}&=\frac{g_3(\alpha_{T_2}-1) f(\Tcal_0)}{2\tilde{\lambda}[ 1+g_3 f(\Tcal_0)+\tilde{\lambda}/(\alpha_{T_2}-1)]}
\nn
&\qquad \qquad \times
\left[ \left(\frac{3 M_{\rm pl}^2(\alpha_{T_2}-1)}{2\tilde{\lambda}^2(\alpha_{T_2}-1+\tilde{\lambda})}\right)^{1/2} \frac{d \tilde{\lambda}}{d t_E}
-\left( \frac{6(\alpha_{T_2}-1+\tilde{\lambda})}{\alpha_{T_2}-1} \right)^{1/2} M_{\rm pl} H_E \right]^2\,.
\end{align}
To obtain a simple expression of the minisuperspace action, we have introduced new parameters $(g_1,g_2,g_3)$ via the relations
\begin{align}
g_1&=\frac{1}{2\alpha_R}\left[\frac{\alpha_X}{\alpha_{T_2}-1}+3\alpha_X+2(\alpha_R-\alpha_C+c_2)+8c_1 \right]\,,
\\
g_2&=\frac{3}{2}-\frac{\alpha_{T_3}-1}{2(\alpha_{T_2}-1)}
-\frac{1}{3\alpha_{\Xcal}} (3\alpha_C-6c_2+8d_1)
+\frac{\alpha_R}{2\alpha_{\Xcal}}(g_1-1)^2\,,
\\
g_3&=\frac{1}{\alpha_{\Xcal}}(3\alpha_{\Xcal}-\alpha_C+2c_2)\,.
\end{align}
As far as $\alpha_{R} \neq 0$, $\alpha_{\Xcal} \neq 0$ and $\alpha_{T_2}\neq 1$, there is a one-to-one correspondence between $(g_1,g_2,g_3)$ and $(c_1,c_2,d_1)$.

The field space metric $G_{ab}d\tilde{\chi}^a d \tilde{\chi}^b $ is a two-dimensional hyperbolic space of which scalar curvature is
\begin{align}
\mathcal{R}=-\frac{1}{3M_{\rm pl}^2}\,.
\end{align}
We define the normalized fields $\chi,\theta$ by
\begin{align}
\tilde{\lambda}&=(\alpha_{T_2}-1) {\rm csch}^{2} \left[\frac{\chi}{\sqrt{6} M_{\rm pl}} \right]\,, \\
\Tcal_0&=2\sqrt{\frac{\alpha_{T_2}-1}{\alpha_{\Xcal}}}M_* \tan \frac{\theta}{2}\,,
\end{align}
under the restrictions $\chi>0, -\pi<\theta<\pi$, where we have used \eqref{alphaT2}.
The kinetic terms are then given by
\begin{align}
\frac{1}{2} G_{ab} \frac{d\tilde{\chi}^a}{dt_E}  \frac{d \tilde{\chi}^b}{dt_E} 
=\frac{1}{2}\left(\frac{d\chi}{dt_E} \right)^2+\frac{1}{2} F^2(\chi) \left(\frac{d\theta}{dt_E}\right)^2\,, \label{kinetic_term}
\end{align}
where
\begin{align}
F(\chi) = \sqrt{\frac{3}{2}}M_{\rm pl} \sinh \left[ \frac{\chi}{\sqrt{6} M_{\rm pl}} \right]\,.
\end{align}

The potential $V_{\rm FLRW}$ can be divided into
\begin{align}
V_{\rm FLRW}=V(\chi)+U(\chi,\theta)\,,
\end{align}
where $V(\chi)$ takes the form \eqref{potential_V} and $U(\chi,\theta)$ is the remaining part. By using the expression
\begin{align}
f(\Tcal_0 )&=2\sin^2 \frac{\theta}{2}= 1-\cos\theta\,,
\end{align}
the remaining part $U$ is written as
\begin{align}
U(\chi,\theta) = \mu_1 (\chi)\left( 1-\cos \theta \right) + \mu_2 (\chi) \left( 1-\cos2\theta \right)\,,
\end{align}
with
\begin{align}
\mu_1(\chi) &=\frac{3M_{\rm pl}^2M_*^2}{2\alpha_{\Xcal}}  \sinh^2 \left[ \frac{\chi}{\sqrt{6} M_{\rm pl}} \right] \left\{
g_1 + \sinh^2 \left[ \frac{\chi}{\sqrt{6} M_{\rm pl}} \right] \left( \frac{\alpha_{T_3}-g_1}{\alpha_{T_2}-1} +2g_2 \right)   \right\}
\nn
&=\frac{F^2 M_*^2}{\alpha_{\Xcal}}\left[ g_1+(y-1)\left(\frac{\alpha_{T_3}-g_1}{2(\alpha_{T_2}-1)}+g_2\right)\right]\,,
\\
\mu_2(\chi) &= - \frac{3 g_2 M_{\rm pl}^2M_*^2 }{4\alpha_{\Xcal}}  \sinh^4 \left[ \frac{\chi}{\sqrt{6} M_{\rm pl}} \right]
\nn
&=-\frac{g_2 F^2 M_*^2}{4\alpha_{\Xcal}}(y-1)\,,
\end{align}
where $y$ is defined by \eqref{def_y}. The non-minimal coupling term takes the form
\begin{align}
\mathcal{L}_{\rm non}=\frac{g_3(y-1)(1-\cos\theta)}{2[y+1+g_3(y-1)(1-\cos \theta)]}\left( \sqrt{\frac{1}{2}(y-1)} \frac{d\chi}{dt_E} +\sqrt{3(y+1)}M_{\rm pl}H_E \right)^2\,.
\end{align}
When $g_2=0$, the potential $U$ is of the same form as the standard axion potential but the scale of the potential $\mu_1$ is ``running'', i.e.~depends on the value of $\chi$. On the other hand, in the generic case with $g_2\neq 0$, the potential consists of two sinusoidal functions with the different ``running'' scales $\mu_1$ and $\mu_2$. Note that not only the scales $\mu_1,\mu_2$ but also ``the axion decay constant'' $F$ depend on the value of $\chi$. The inflationary model with two or more than two sinusoidal functions is called the multi-natural inflation~\cite{Czerny:2014wza}. In the direction to $\theta$, the potential is similar to the multi-natural inflation potential with the running scales $\mu_1,\mu_2$ and the running decay constant $F$.

As a result, after the field redefinitions, the minisuperspace action is given by a non-linear sigma model with the hyperbolic field space coupled to GR and the remaining non-minimal coupling $\mathcal{L}_{\rm non}$. The potential is the sum of the Starobinsky-like potential $V$ and the multi-natural inflation-like potential $U$. In particular, the non-minimal coupling disappears if $g_3=0$. We shall refer to the frame of the action \eqref{Einstein_frame} as the quasi-Einstein frame and to the original one \eqref{L_FLRW_C} as the Jordan frame of the minisuperspace action, respectively. Note that even in the $g_3=0$ case we should not call \eqref{Einstein_frame} the Einstein frame in a strict sense since this description is valid only for the homogeneous and isotropic background dynamics and does not hold for perturbations.

The function $A$ defined by \eqref{conformal_A} diverges as either $\tilde{\lambda}\rightarrow \infty$ or $\Tcal_0 \rightarrow \pm \infty$. The points $\tilde{\lambda}\rightarrow \infty$ or $\Tcal_0 \rightarrow \pm \infty$ are the Big Bang singularity of the Jordan frame, $a\rightarrow 0$. Regarding the normalized fields  these singular points correspond to $\chi\rightarrow 0$ ($y\rightarrow 1$) and $\theta \rightarrow \pm \pi$, respectively. However, the singularities cannot be seen in the quasi-Einstein frame action and the singularities are just local extrema of the potential in the direction to $\chi$ or $\theta$, respectively (see the comments after \eqref{potential_V}).

\subsection{Test field approximation of $\theta$}
We now study the case when the universe is dominated by $\chi$; that is, the background dynamics is effectively described by \eqref{Einstein_single}. Since the background dynamics of \eqref{Einstein_single} was discussed in the previous section, we only study the dynamics of $\theta$ which may be treated as a test field. 
During the slow roll regime, the non-minimal coupling term can be approximated as
\begin{align}
\mathcal{L}_{\rm non}
&\simeq \frac{g_3(y+1)(y-1)(1-\cos\theta)}{2[y+1+g_3(y-1)(1-\cos \theta)]} 3M_{\rm pl}^2 H_E^2
\nn
&\simeq \frac{g_3 F^2 M_*^2(y+1-2\alpha_{T_2})^2(y+1)(1-\cos\theta)}{8\alpha_R(\alpha_{T_2}-1)^2[y+1+g_3(y-1)(1-\cos \theta)]} \,,
\end{align}
where we have ignored the slow-roll suppressed term $d\chi/dt_E$ in the first line and then used the Einstein equation $3M_{\rm pl}^2H_E^2\simeq V$ to obtain the second line. The dynamics of $\theta$ is thus effectively described by the action
\begin{align}
S_{\theta,{\rm eff}}&=\int dt_E d^3\bm{x} a_E^3 \left[ \frac{F^2}{2}\left(\frac{d\theta}{dt_E}\right)^2-U_{\rm eff} \right]\,, \\
U_{\rm eff}&=U-\frac{g_3 F^2 M_*^2(y+1-2\alpha_{T_2})^2(y+1)(1-\cos\theta)}{8\alpha_R(\alpha_{T_2}-1)^2[y+1+g_3(y-1)(1-\cos \theta)]} \,.
\end{align}
The test field approximation may be valid as long as
\begin{align}
 \frac{F^2}{2}\left(\frac{d\theta}{dt_E}\right)^2 \ll M_{\rm pl}^2 H_E^2
\,, \quad
|U_{\rm eff}| \ll M_{\rm pl}^2 H_E^2
\,.
\end{align}

We first read the effective mass of $\theta$ around $\theta=0$. The effective potential is
\begin{align}
U_{\rm eff}\simeq \frac{F^2M_*^2}{2} \left[ \frac{(y-1)\alpha_{T_3} }{2\alpha_{\Xcal} (\alpha_{T_2}-1)} +\frac{g_1 (2\alpha_{T_2}-y-1)}{2\alpha_{\Xcal}(\alpha_{T_2}-1)} -\frac{g_3  (2\alpha_{T_2}-y-1)^2}{8\alpha_R (\alpha_{T_2}-1)^2} \right] \theta^2 +O(\theta^4)
\,.
\end{align}
During the inflationary regime, the analytic results \eqref{y_top} and \eqref{y_slope} conclude $y-1 \ll \alpha_{T_2}$ for $N_e \gg 1$; then, the squared effective mass is
\begin{align}
m_{\theta,{\rm eff}}^2=M_*^2 \left[ \frac{g_1}{\alpha_{\Xcal}}-\frac{g_3}{2\alpha_R} \right]
=\frac{M_*^2}{2\alpha_R}\left[ \frac{1}{\alpha_{T_2}-1}+\frac{2\alpha_R-\alpha_C+8c_1}{\alpha_{\Xcal}} \right]\,, \label{eff_mass}
\end{align}
which can be either positive or negative even if the squared mass of $0^-$ around the Minkowski background, $m_{0^-}^2=\alpha_{T_3}M_*^2/\alpha_{\Xcal}$, is positive. 
If the dimensionless constants are of order unity, the effective mass is the same order of magnitude of the Hubble scale $H_E \sim M_*$.

Due to the factor $F=\sqrt{\frac{3(y-1)}{4}}M_{\rm pl}$, the test field approximation can be used when $\chi$ is near the top of the potential ($y\simeq 1$). The effective potential $U_{\rm eff}$ is approximated as
\begin{align}
U_{\rm eff}\simeq F^2 m_{\theta,{\rm eff}}^2 (1-\cos \theta)
\,.
\end{align}
The point $\theta=0~(\Tcal_0=0)$ is the minimum of the potential for $m_{\theta,{\rm eff}}^2 >0$ while the singular points $\theta=\pm \pi~(\Tcal_0 \rightarrow \pm \infty)$ are the minima when $m_{\theta,{\rm eff}}^2< 0$. The field $\theta$ tends to approach the minimum $\theta=0$ or $\theta=\pm \pi$ depending on the sign of $m_{\theta,{\rm eff}}^2$. A special case is  $m_{\theta,{\rm eff}}^2 = 0$ where the effective mass of $\theta$ vanishes in the limit $\chi \rightarrow 0$.
On the slope of the potential $V$, namely for $y-1\gtrsim O(1)$, $\theta$ may be treated as the test field if $\theta \ll 1$ and if $(d\theta/d t_E)^2\ll H_E^2$ but, in general, the back-reaction of $\theta$ can not be ignored.

In the case of $m_{\theta,{\rm eff}}^2 <0$, the field $\theta$ must have an expectation value during the inflationary regime independently of the initial condition. The trajectory of the fields $\chi$ and $\theta$ will be shown in the next subsection. On the other hand, the value of $\theta$ should decrease as the universe expands when $m_{\theta,{\rm eff}}^2 >0$. The typical value of $\theta$ at the horizon crossing must depend on the initial condition of the inflation and the effective mass.

We estimate the typical value of $\theta$ for $m_{\theta,{\rm eff}}^2 >0$ in the remaining part of this subsection.
Let us consider the parameter space $m_{\theta,{\rm eff}}^2 >0$ and $\alpha_{T_2} \gg 1$, and suppose that the universe was born of a point close to the singularity $\chi \simeq 0,|\theta|=O(1)$. We denote the e-folds of the initial time as $N_i$.
The theory \eqref{LG} must be valid only after the Jordan frame Hubble expansion rate,
\begin{align}
H=\frac{a'}{aN}=\sqrt{\frac{\alpha_{T_2}-1}{(y-1)^2(1+\cos \theta)}} 
\left[ \sqrt{y-1}\left(2H_E-\frac{d\theta}{dt_E} \tan \frac{\theta}{2} \right) +\sqrt{\frac{2}{3}(y+1)} \frac{1}{M_{\rm pl}} \frac{d\chi}{dt_E} \right]\,,
\end{align} 
becomes below the Planck scale. We have
\begin{align}
 \frac{d\chi}{dt_E} \simeq -\frac{dV/d\chi}{3H_E} \simeq  \sqrt{2\epsilon_V }M_{\rm pl} H_E 
\,,
\end{align}
under the test field approximation of $\theta$ as well as the slow roll approximation of $\chi$. Therefore, near the singularity $\chi\simeq 0~(y\simeq 1)$, the Jordan frame Hubble expansion rate is
\begin{align}
H \sim \sqrt{\frac{\alpha_{T_2}}{y-1}} H_E \,,
\end{align}
as far as $\left| \frac{d\theta}{dt_E} \tan \frac{\theta}{2} \right| \lesssim H_E$. If the universe was born at the Planck scale $H \sim M_{\rm pl}$, the e-folds of the initial time $N_i$ is given by
\begin{align} 
N_i \sim \frac{3}{2} \alpha_{T_2} \ln \frac{M_{\rm pl}}{\alpha_{T_2}^{1/2} H_E} \,.
\end{align}
where we used \eqref{y_top} which can be applied when $N_i \gg \alpha_{T_2}$. The typical value is $N_i \sim 10 \alpha_{T_2}$ in the validity range of the approximation; for instance, the values $\alpha_{T_2}\sim 100, M_{\rm pl}/H_E \sim 10^5$ lead to $N_i/\alpha_{T_2}\sim 10$ for which $N_i \gg \alpha_{T_2}$ is barely satisfied.
To estimate the typical value of $\theta$ at the horizon crossing of the CMB scale, we approximate the effective potential of $\theta$ as the quadratic potential with the mass \eqref{eff_mass},
\begin{align}
U_{\rm eff} \simeq \frac{1}{2}F^2 m_{\theta, {\rm eff}}^2 \theta^2
\,.
\end{align}
The approximate solution of $\theta$ is
\begin{align}
\theta \propto a^{-\frac{3}{2} \pm \sqrt{\frac{9}{4}-\frac{m_{\theta,{\rm eff}}}{H_E^2}} }\,,
\end{align}
where we have used
\begin{align}
\frac{d\ln F}{dt_E}=\sqrt{\frac{y+1}{6(y-1)}} \frac{1}{M_{\rm pl}}\frac{d\chi}{dt_E} \simeq \frac{2(y+1)}{3(y+1-2\alpha_{T_2})}H_E \ll H_E
\end{align}
for $\alpha_{T_2} \gg 1$. The typical value of $\theta$ at the horizon crossing $(N_e \simeq 50 \ll N_i)$ is thus
\begin{align}
|\theta_{N_e}| \sim \exp\left[ \left(-\frac{3}{2} + \sqrt{\frac{9}{4}-\frac{m_{\theta,{\rm eff}}^2 }{H_E^2}} \right) N_i \right]\,,
\end{align}
for $m_{\theta, {\rm eff}}^2 < \frac{9}{4}H_E^2$ and
\begin{align}
|\theta_{N_e}| \sim \exp \left[ -\frac{3}{2}N_i \right]\,,
\end{align}
for $m_{\theta, {\rm eff}}^2 > \frac{9}{4}H_E^2$, respectively. When the absolute value of the exponent is much larger than unity, we may ignore the background value of $\theta$ in the observable range of the universe since the amplitude of $\theta$ exponentially decays. The effective mass is given by \eqref{eff_mass} and the Hubble rate can be expressed by the parameters of the theory by using the Friedmann equation,
\begin{align}
\frac{m_{\theta,{\rm eff}}^2}{H_E^2} \sim \frac{\alpha_R}{\alpha_{\Xcal}}g_1-\frac{g_3}{2} = \frac{1}{2(\alpha_{T_2}-1)}+\frac{2\alpha_R-\alpha_C+8c_1}{2\alpha_{\Xcal}}\,.
\end{align}
 We may classify the scenarios into three cases
\begin{align}
 {\rm case~i}:~ \frac{1}{2(\alpha_{T_2}-1)}+\frac{2\alpha_R-\alpha_C+8c_1}{2\alpha_{\Xcal}} &\gg \frac{1}{10\alpha_{T_2}}
\,, \label{caseI} \\
 {\rm case~ii}:~ \frac{1}{2(\alpha_{T_2}-1)}+\frac{2\alpha_R-\alpha_C+8c_1}{2\alpha_{\Xcal}} &\sim \frac{1}{10\alpha_{T_2}}
\,, \label{caseII} \\
 {\rm case~iii}:~ \frac{1}{2(\alpha_{T_2}-1)}+\frac{2\alpha_R-\alpha_C+8c_1}{2\alpha_{\Xcal}} &\ll \frac{1}{10\alpha_{T_2}}
\,. \label{caseIII}
\end{align}
In the case i, the background value of $\theta$ can be ignored due to the exponential suppression and the test field approximation of $\theta$ is trivially justified. In the case ii and the case iii, however, $\theta$ cannot be ignored. In the case ii, the value of $\theta$ must barely decrease up until the horizon crossing. The test field approximation may be valid on the slope as the leading approximation and the effect of $\theta$ can be included perturbatively. In the case iii, on the other hand, we cannot use the test field approximation during the inflationary regime.

\subsection{Trajectories of $\chi$ and $\theta$}
We then study general trajectories of $\chi$ and $\theta$ without approximations.
We specially focus on the dynamics of $\chi$ and $\theta$ before the e-folds $N_e=50$ in order to discuss the initial condition dependence of the inflation.
Needless to say, the precise dynamics of the universe at a scale close to the Planck scale cannot be discussed without knowledge of quantum gravity.
We, however, suppose that the theory \eqref{LG} is valid at a scale barely smaller than the Planck scale and set initial conditions at this scale.

\begin{figure}[t]
\centering
  \includegraphics[width=5cm,angle=0,clip]{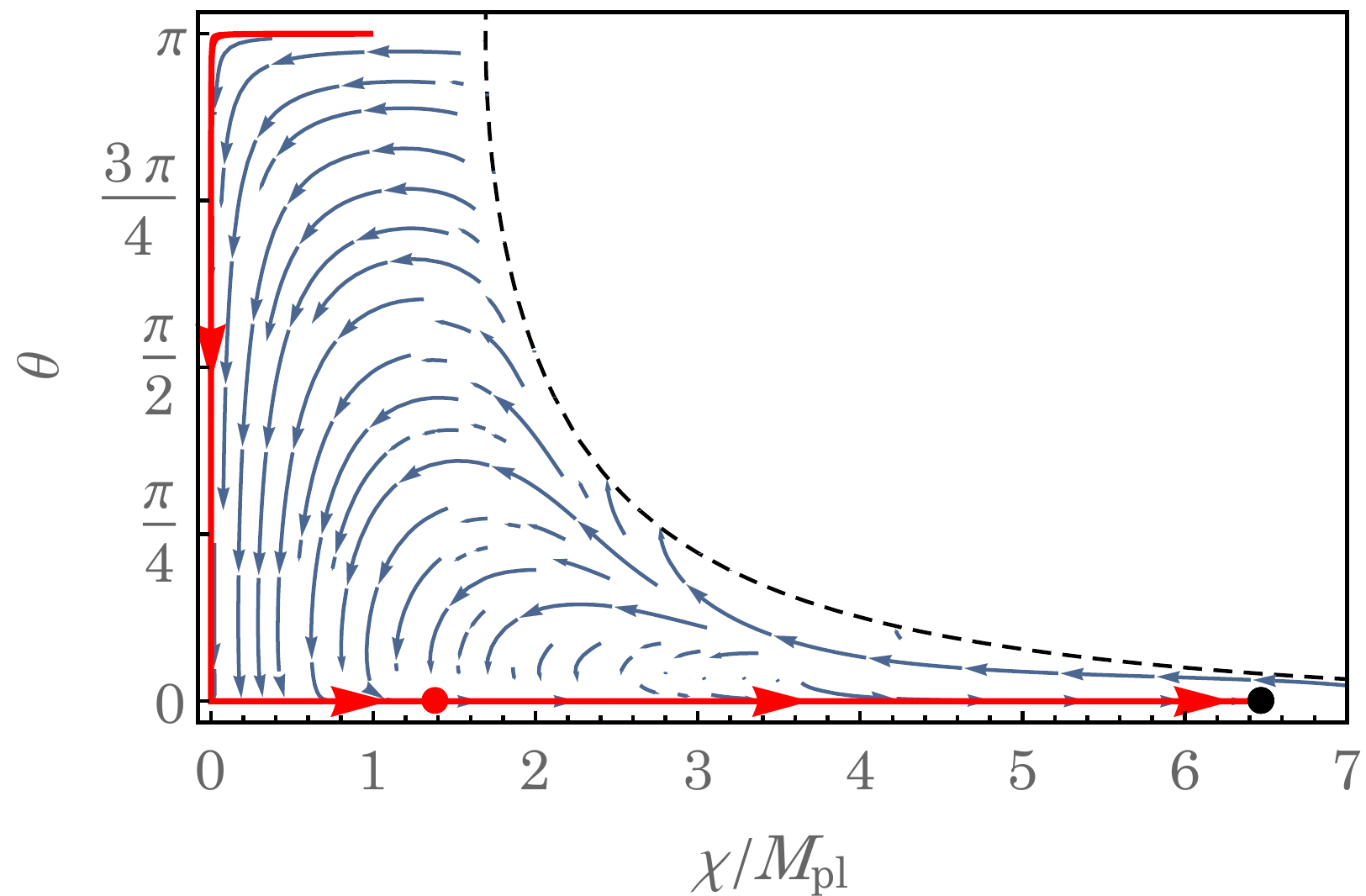}
  \includegraphics[width=5cm,angle=0,clip]{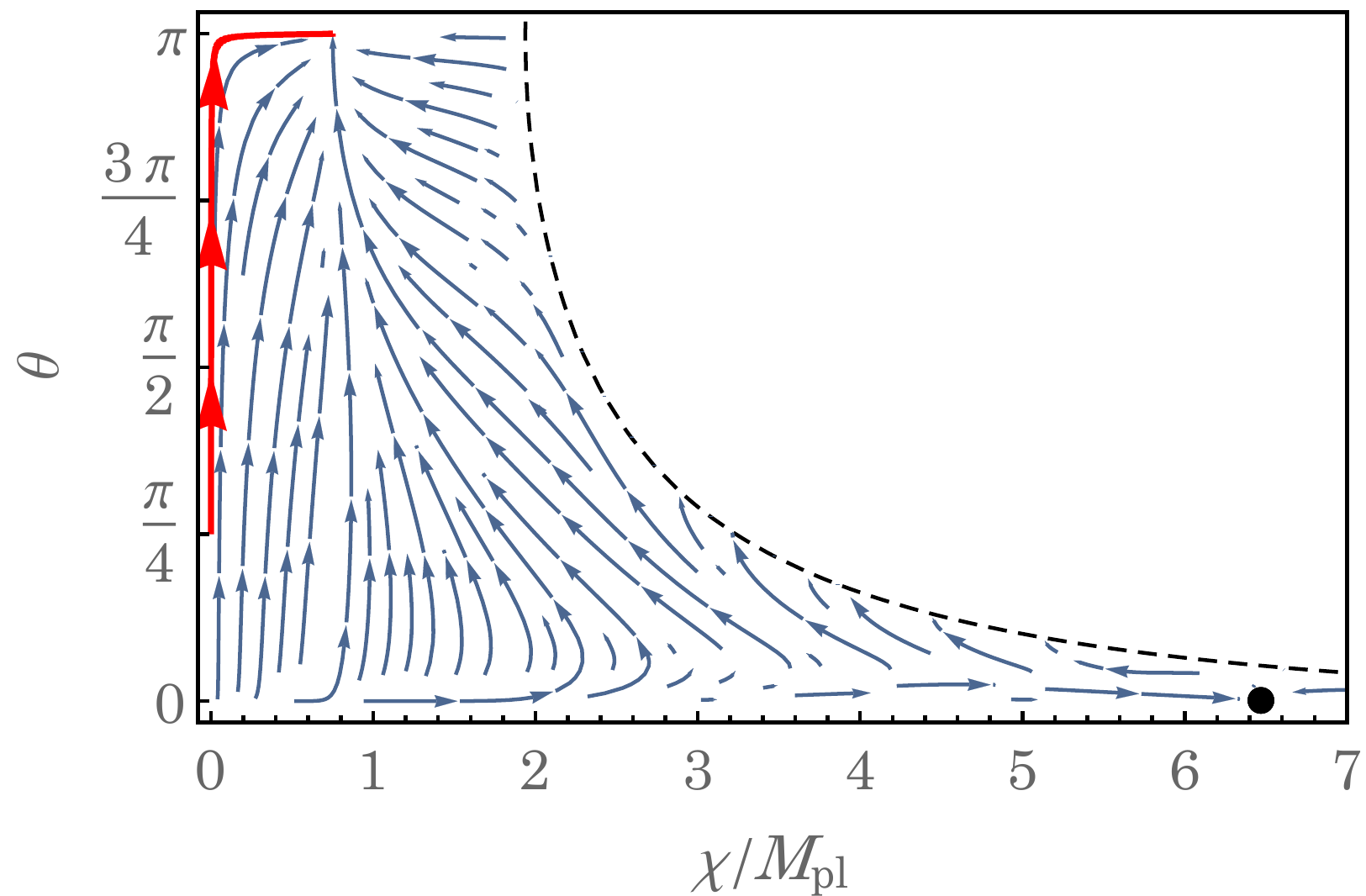}
  \includegraphics[width=5cm,angle=0,clip]{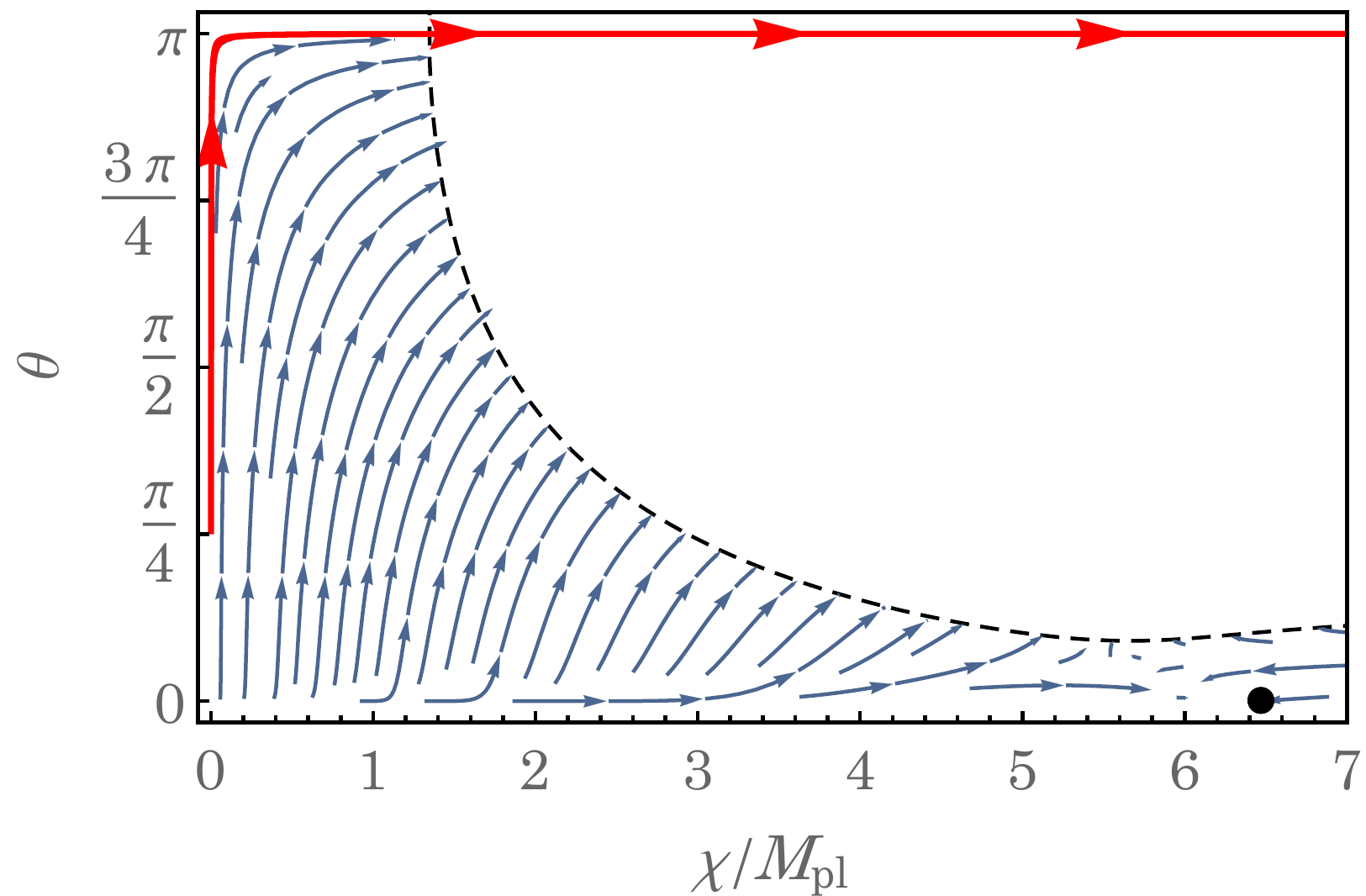}
  \\
\qquad
Model I
\qquad\qquad \qquad \qquad \quad
Model II
\qquad \qquad \qquad \qquad \quad
Model III
\caption{Blue arrows show the acceleration vector $( d^2\chi/dt_E^2, d^2\theta/dt_E^2 )$ on the two-dimensional subspace of the four-dimensional space defined by $d\chi/dt_E=d\theta/dt_E=0$. The red curves are the projections of the trajectories of numerical solutions. The lines $\chi=0$ and $\theta=\pi$ are the Big Bang singularity of the Jordan frame. The black dot is the potential minimum $\chi=\chi_M=6.48M_{\rm pl},\theta=0$ and the red dot is the point at the e-folds $N_e=50$ of the numerical solution. The black dashed curves represent $H_E=0$ (but $dH_E/dt_E \ne 0$ there). We set $\alpha_{T_2}=50, \alpha_{T_3}=1, M_*/M_{\rm pl}=10^{-5}, \alpha_R=1,c_1=c_2=d_1=0$ with $\alpha_{\Xcal}=2, \alpha_C=1$ (left), $\alpha_{\Xcal}=1, \alpha_C=3$ (middle), and $\alpha_{\Xcal}=1,\alpha_C=3$ (right), which correspond to $m_{\theta,{\rm eff}}^2 =0.26 M_*^2$, $m_{\theta,{\rm eff}}^2=-0.24M_*^2$, and  $m_{\theta,{\rm eff}}^2=-0.49M_*^2$, respectively.  }
\label{fig_PS_alphaC}
\end{figure}

\begin{figure}[t]
\centering
\includegraphics[width=5cm,angle=0,clip]{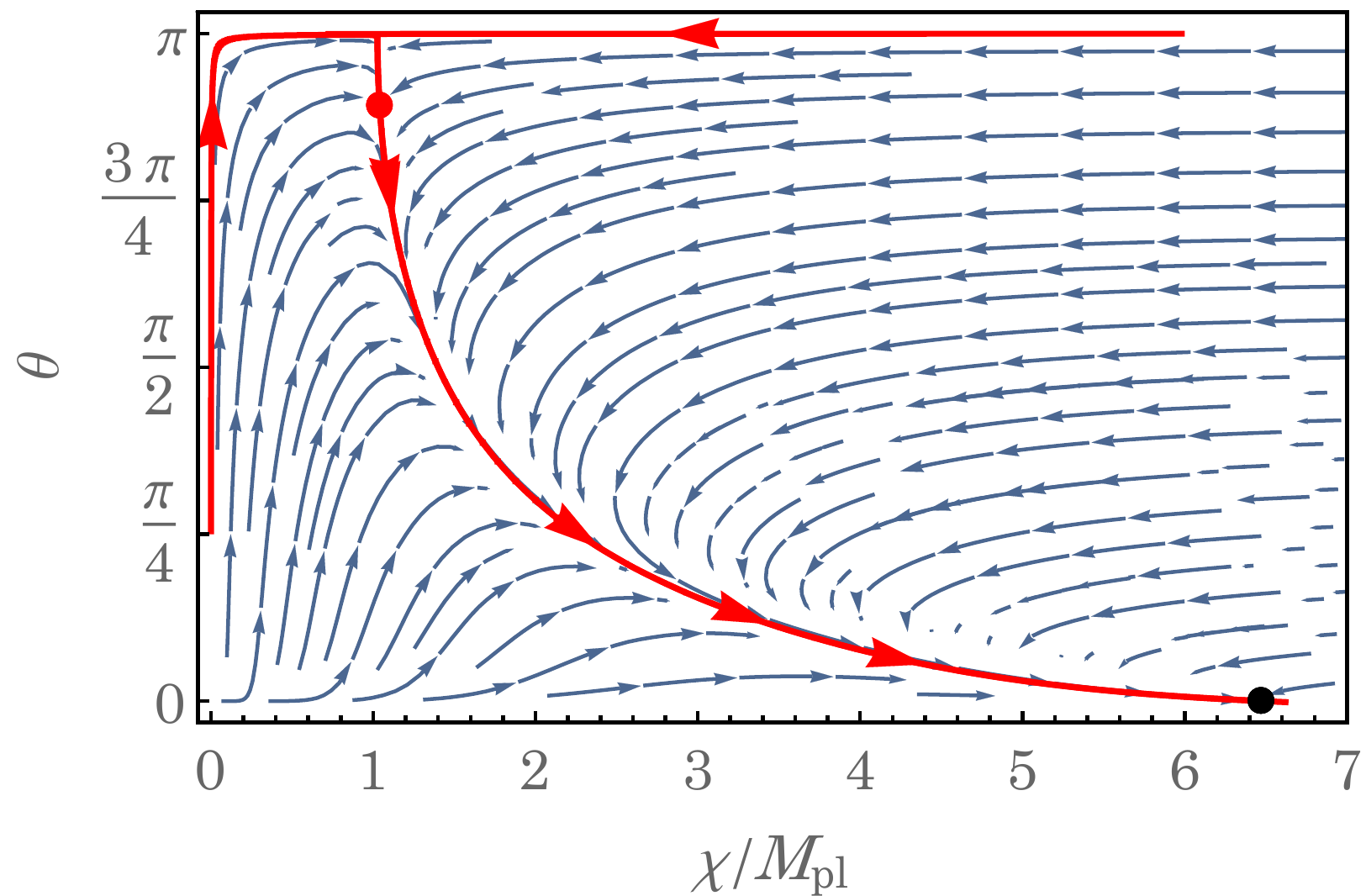}
\includegraphics[width=5cm,angle=0,clip]{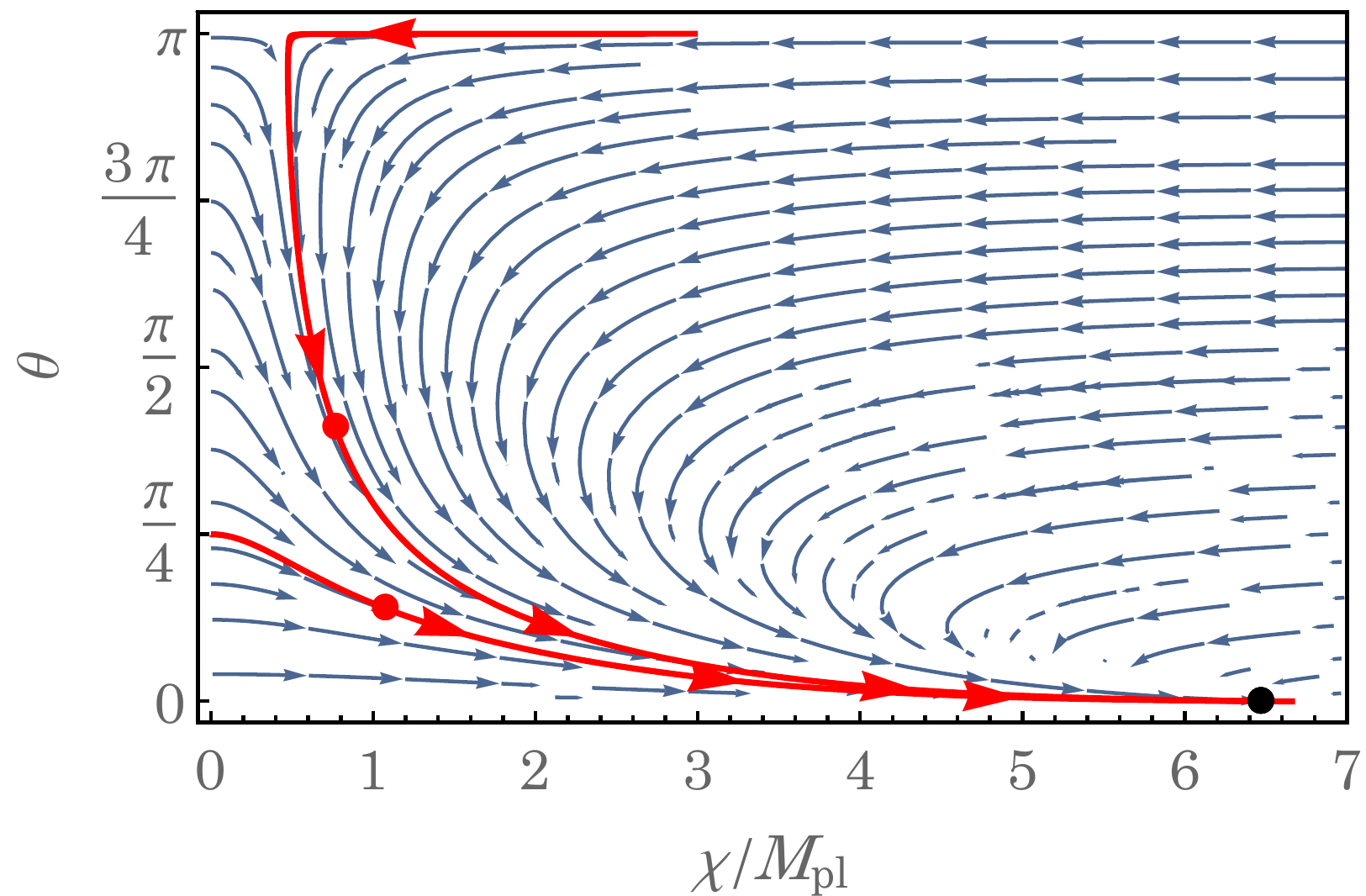}
\includegraphics[width=5cm,angle=0,clip]{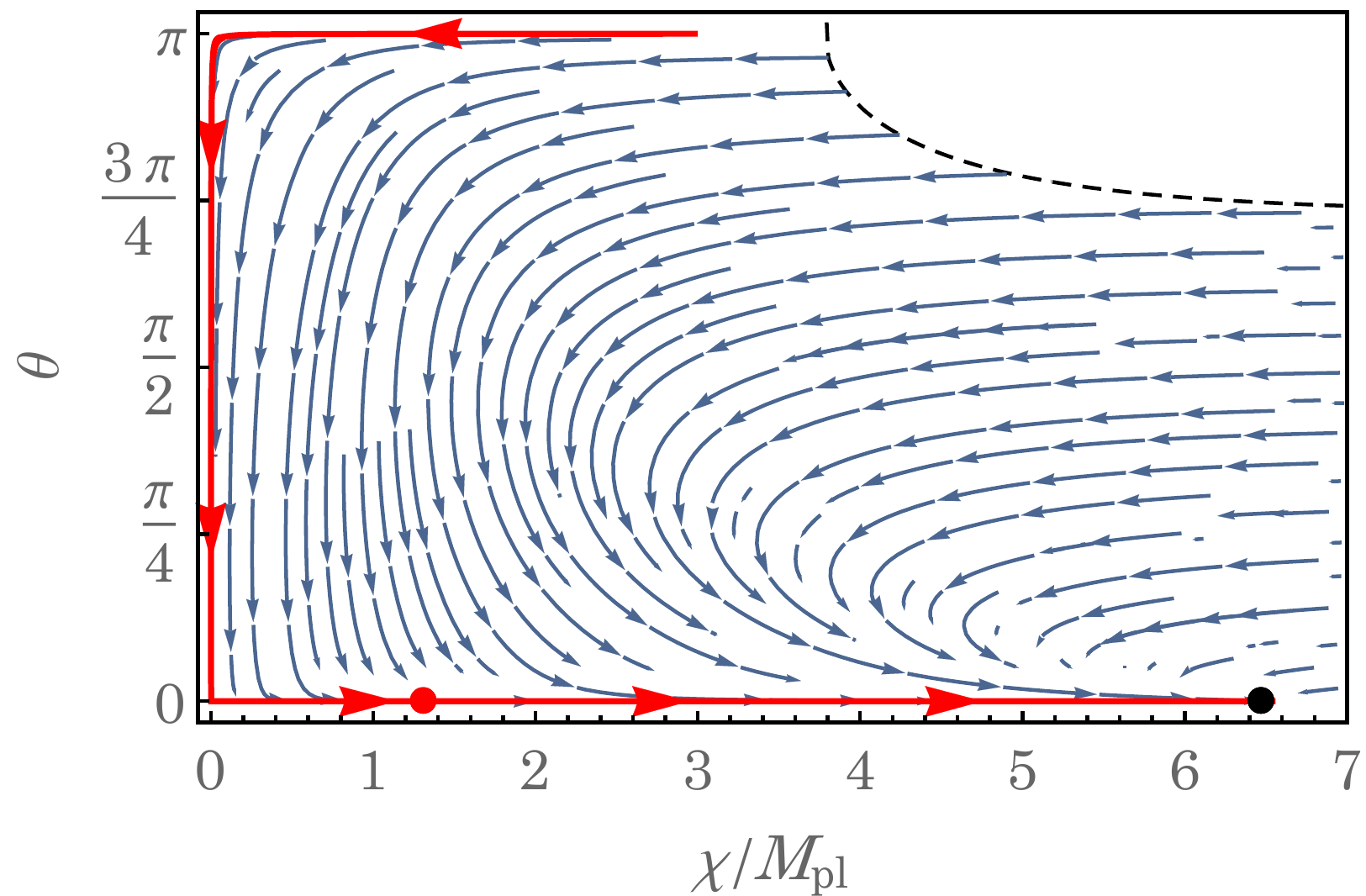}
\\
\qquad
Model IV
\qquad\qquad \qquad \qquad \quad
Model V
\qquad \qquad \qquad \qquad \quad
Model VI
\caption{The same figures as Fig.~\ref{fig_PS_alphaC} where $\alpha_{T_2}=50, \alpha_{T_3}=1, M_*/M_{\rm pl}=10^{-5}, \alpha_R=1, \alpha_{\Xcal}=1,g_1=-1/5,g_2=1/5$ with $g_3=-1/5$ (left), $g_3=-2/5$ (middle), and $g_3=-3/5$ (right), which correspond to $m_{\theta,{\rm eff}}^2 =- M_*^2/10$, $m_{\theta,{\rm eff}}^2=0$, and $m_{\theta,{\rm eff}}^2=M_*^2/10$, respectively.  }
\label{fig_PS_g3}
\end{figure}

The overall behavior of the inflationary trajectories of the fields may be traced by considering the two-dimensional subspace of the four-dimensional phase space defined by $d\chi/dt_E=d\theta/dt_E=0$ since the fields should be slowly rolling during the inflation. On the two-dimensional subspace the velocity vector $( d\chi/dt_E, d\theta/dt_E )$ vanishes. We thus plot the acceleration vector $( d^2\chi/dt_E^2, d^2\theta/dt_E^2 )$ on the two-dimensional subspace in Figs~\ref{fig_PS_alphaC} and \ref{fig_PS_g3} as blue arrows. In these figures, the plots are shown only in the range $0\leq \theta <\pi$ since the minisuperspace action is invariant under $\theta \rightarrow -\theta $ due to the parity invariance. The black dashed curves show points with $H_E=0$ but do not represent the end of inflation even in the Einstein frame since $dH_E/dt_E \ne 0$ there. The black dot shows the potential minimum at $(\chi,\theta) = (\chi_M,0)$ and corresponds to the end of reheating after inflation. In the same figures, numerical solutions with initial conditions $d\chi/dt_E=d\theta/dt_E=0$ and $H \sim 0.1M_{\rm pl}$ are shown by red curves\footnote{For numerical calculations, we solve the evolution equations of $\chi,\theta$ and $H_E$. The Friedmann equation is used to set the initial condition and used to check the accuracy of the calculations.}. In the cases where the fields evolve towards the potential minimum, we also plot a red dot which represents the e-folds $N_e= 50$, where the end of inflation is determined by $-(1/H_E^2)(dH_E/dt_E)=1$.

If the absolute value of $m_{\theta,{\rm eff}}^2$ is large enough (but still of $\mathcal{O}(M_*^2)$) then the graceful exit from inflation towards reheating tends to favor positive $m_{\theta,{\rm eff}}^2$. To illustrate this point, Fig~\ref{fig_PS_alphaC} highlights three models with different values of $m_{\theta,{\rm eff}}^2$: $m_{\theta,{\rm eff}}^2 =0.26 M_*^2$ (Model I), $m_{\theta,{\rm eff}}^2=-0.24M_*^2$ (Model II) and  $m_{\theta,{\rm eff}}^2=-0.49M_*^2$ (Model III). FIG~\ref{fig_PS_g3} focuses on more subtle dependence of the behavior of the system on the value of $m_{\theta,{\rm eff}}^2$ by showing three models with smaller absolute values of $m_{\theta,{\rm eff}}^2$: $m_{\theta,{\rm eff}}^2 =- M_*^2/10$ (Model IV), $m_{\theta,{\rm eff}}^2=0$ (Model V) and $m_{\theta,{\rm eff}}^2=M_*^2/10$ (Model VI).

When $m_{\theta, {\rm eff}}^2>0$ (Models I and VI), the fields $(\chi,\theta)$ first tend to approach $\chi \simeq 0, \theta \simeq 0$ from the Big Bang singularity $\chi \simeq 0$ or $\theta \simeq  \pi$; then, the fields roll down toward the potential minimum $\chi=\chi_M,\theta=0$ along $\theta \simeq 0$. The Starobinsky-like inflationary scenario discussed in \S.~\ref{sec_heavy} may be naturally realized from the Big Bang singularity. 
On the other hand, in the negative squared mass case, $m_{\theta, {\rm eff}}^2<0$, the fields first go to $\chi \neq  0 $ and $\theta \simeq  \pi$. In Model IV, the fields then roll down toward the potential minimum along the curve from $\chi\simeq M_{\rm pl} , \theta \simeq \pi $. On the other hand, in Models II and III, the fields do not approach the potential minimum. In the vanishing effective mass case (Model V), the trajectory depends on the initial condition. If the initial condition is $\chi \simeq 0$, the fields directly move toward the potential minimum whereas the trajectory is similar to Model IV if the initial condition is $\theta \simeq \pi$. In these examples, Models II and III do not provide a realistic inflationary scenario since the inflation does not end while other models can provide a graceful exit from inflation.

\begin{figure}[t]
\centering
\includegraphics[width=5.8cm,angle=0,clip]{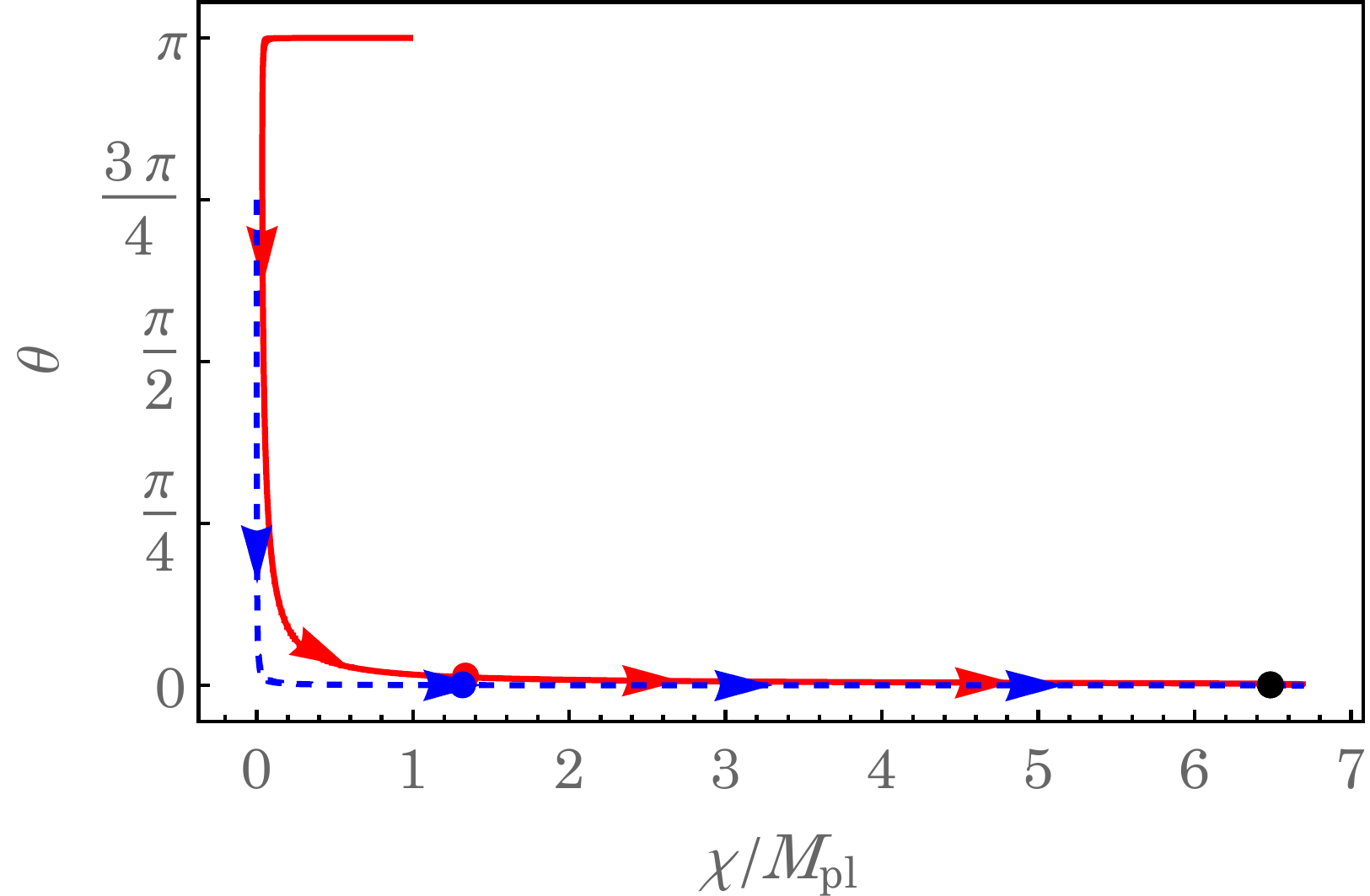}
\quad
\includegraphics[width=5.8cm,angle=0,clip]{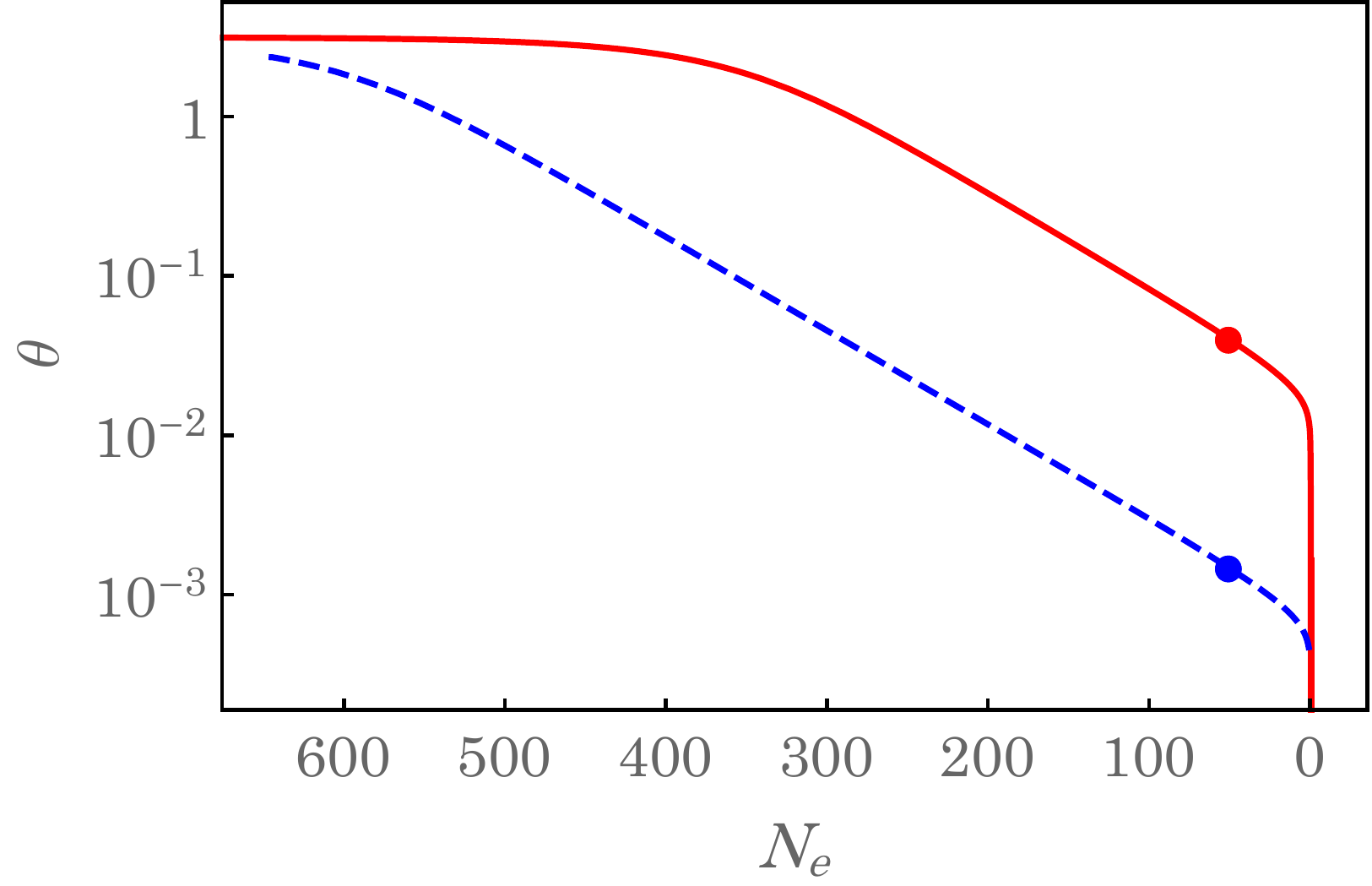}
\caption{The inflationary trajectories of the fields on the $(\chi, \theta)$ plane (left panel) and the evolutions of $\theta$ (right panel) in the model $\alpha_{T_2}=50, \alpha_{T_3}=1, M_*/M_{\rm pl}=10^{-5}, \alpha_R=1, \alpha_{\Xcal}=2,\alpha_C=2 $ and $c_1=c_2=d_1=0$. The initial conditions are $d\chi/dt_E=d\theta/dt_E=0$ with $\chi=10^{-3}M_{\rm pl},\theta=3\pi/4$ (red) and $\chi=M_{\rm pl},\theta=3.999\pi/4$ (blue), which correspond to $H \simeq 0.22M_{\rm pl}$ and $H=0.38M_{\rm pl}$, respectively, at the initial time. The dots on the curves are the points at the e-folds $N_e=50$.
}
\label{fig_caseII}
\end{figure}

In Model I where the effective mass of $\theta$ is comparable to the inflationary Hubble scale and then is classified into the case i \eqref{caseI}, the value of $\theta$ at the e-folds $N_e=50$ is extremely small (the numerical solution gives its value as $\theta \simeq 10^{-105}$ where the initial condition is $\chi=M_{\rm pl},\theta=3.999\pi/4$ and $d\chi/dt_E=d\theta/dt_E=0$ which gives $H \simeq 0.38M_{\rm pl}$ at the initial time). On the other hand, the value of $\theta$ is expected to be not so small in the case ii \eqref{caseII}. Numerical solutions in the case ii are shown in Fig.~\ref{fig_caseII} where we assume
\begin{align}
2\alpha_R-\alpha_C+8c_1 =0
\,,
\end{align}
to satisfy \eqref{caseII}. At the e-folds $N_e=50$, the values of $\theta$ are $\theta \simeq 0.03$ for the initial condition $\chi=10^{-3}M_{\rm pl},\theta=3\pi/4$ (red curves) and $\theta\simeq 0.001$ for $\chi=M_{\rm pl},\theta=3.999\pi/4$ (blue curves), respectively.

As a result, there are mainly three possible inflationary scenarios depending on the sign of $m_{\theta,{\rm eff}}^2$ and its absolute value. For the case with $m_{\theta,{\rm eff}}^2>0$, the background dynamics of the universe can be approximated by \eqref{Einstein_single} since the value of $\theta$ decays. The value of $\theta$ can be ignored at the horizon crossing if the effective mass is large enough to satisfy the case i condition \eqref{caseI} whereas $\theta$ is not so small for the case ii \eqref{caseII}. It would be worth emphasizing that the fields move toward $\chi\simeq 0,\theta\simeq 0$, namely the top of the Starobinsky-type potential $V$, even if the initial condition is $\chi \neq 0$, as shown for Models I and IV in Figs~\ref{fig_PS_alphaC} and \ref{fig_PS_g3}, respectively. Although the potential form of $\chi$ is similar to the hilltop inflation, $\chi$ can be automatically set in the top of the potential from generic initial conditions. The second possible scenario is the case with $m_{\theta,{\rm eff}}^2<0$ where both fields have non-zero values during the inflation. As in the first scenario, the same trajectory of the fields can be realized from generic initial conditions (Model IV in Fig \ref{fig_PS_g3}). Finally, if the effective mass vanishes or is too small, the trajectory strongly depends of the initial conditions (Model V in Fig \ref{fig_PS_g3}).


\section{Tensor perturbations}
\label{sec_tensor}
We then discuss linear perturbations around the FLRW background. In the present paper, we shall only focus on the tensor perturbations and we leave the scalar and vector perturbations for a future study. As for the background analysis, we have integrated out the auxiliary variable $\varphi,\Xi_{\mu\nu},A_{\mu\nu}$ to obtain a useful expression of the minisuperspace action. On the other hand, the use of the equivalent action \eqref{L_eq} is indeed useful for computations of perturbations.

In the tensor sector of linear perturbations, the irreducible components $T_{\mu},\Tcal_{\mu}$ are unperturbed. By using three tensor perturbation variables
$h^{TT}_{ij},t^{TT}_{ij},\tau^{TT}_{ij}$, the metric and $\Ta_{\mu\nu\rho}$ are given by
\begin{align}
ds^2&=-N^2(\tau)d\tau^2+a^2(\tau)\gamma_{ij} dx^i dx^j\,, \quad \gamma_{ij}=e^{h^{TT}_{ij}}
\,, \\
\Ta^i{}_{j0}&=N \delta^{ik} t^{TT}_{jk}
\,, \quad 
\Ta_{ijk}=a^3 \epsilon_{jkl}\delta^{lm}\tau^{TT}_{im}
\,,
\end{align}
where all tensor perturbation variables are transverse and traceless, e.g.
\begin{align}
\delta^{ij}h^{TT}_{ij}=0\,,\quad \delta^{ij}\partial_i h^{TT}_{jk} =0
\,.
\end{align}
The variables $h^{TT}_{ij}$ and $t^{TT}_{ij}$ are parity even while $\tau^{TT}_{ij}$ is parity odd.
The auxiliary variables are given by
\begin{align}
\lambda &=\bar{\lambda}(\tau) \,, \quad \varphi=\bar{\varphi}(\tau)\,, \quad A_{\mu\nu}=0\,, \nn
\Xi_{00}&=-N^2\bar{\Xi}_{0}(\tau)\,, \quad \Xi_{ij}=\frac{1}{3} \bar{\Xi}(\tau)\gamma_{ij}+a^2 \Xi^{TT}_{ij}\,,
\end{align}
for tensor perturbations, where $\Xi^{TT}_{ij}$ is transverse-traceless. The background parts of the auxiliary variables are determined by the background equation of motion as
\begin{align}
\bar{\lambda}-1&=\frac{\alpha_R}{3}(\hat{\rho}_B-3\hat{p}_B)
\,, \\
\bar{\varphi}&=-\frac{\alpha_{\Xcal}}{M_*^2}\left[ \frac{\Tcal_0'}{N}+\left(3H+\frac{2}{3}T_0\right)\Tcal_0 \right]
\,, \\
\bar{\Xi}_{0}&=-\frac{\alpha_C}{3}(2\hat{\rho}_B+3\hat{p}_B)
\,, \\
\bar{\Xi}&=2\alpha_C \hat{\rho}_B\,,
\end{align}
where
\begin{align}
\hat{\rho}_B&:=-\frac{1}{M_*^2} \GA^0{}_0=\frac{3}{M_*^2}\left[ \left( H+\frac{1}{3}T_0 \right)^2-\frac{1}{4}\Tcal_0^2\right]
, \label{rho_B} \\
\hat{p}_B&:=\frac{1}{3M_*^2 } \sum_{i=1}^3\GA^i{}_i=\frac{1}{M_*^2} \left[-\frac{2(3H'+T_0')}{3N}-\frac{1}{9}(3H+T_0)(9H+ T_0) +\frac{1}{4}\Tcal_0^2 \right]\,,
\end{align}
are the $00$ and spatial components of the background Einstein tensor, respectively.

In the momentum space, the tensor perturbations can be decomposed as
\begin{align}
h^{TT}_{ij}=h_L Y^{L}_{ij}+h_RY^R_{ij}
\,, ~
t^{TT}_{ij}=t_L Y^{L}_{ij}+t_RY^R_{ij}
\,, ~
\Xi^{TT}_{ij}=\Xi_L Y^{L}_{ij}+\Xi_RY^R_{ij}
\,, ~
\tau^{TT}_{ij}=\tau_L Y^{L}_{ij}-\tau_RY^R_{ij}\,,
\end{align}
where the helicity basis $Y^A_{ij},~(A=L,R)$ satisfy
\begin{align}
\partial^2 Y^{A}_{ij}=-k^2 Y^{A}_{ij}\,, \quad \partial_i Y^{Ai}{}_j=0\,, \quad Y^{Ai}{}_i=0\,, \quad
\epsilon^{(i|jk}\partial_j Y^{A}{}_k{}^{l)}= \epsilon_A k Y^{Ail}\,,
\end{align}
with $\epsilon_L=- 1,\epsilon_R=+ 1$. The minus sign in front of $\tau_R$ has been inserted in order that the L sector and the R sector obey the same equations if the background preserves parity.

By the use of the equivalent form \eqref{L_eq}, the variable $t_A$ is non-dynamical in the quadratic Lagrangian of the tensor perturbations and thus can be integrated out. Then, we obtain the quadratic Lagrangian in terms of three variables $h_A,\Xi_A$ and $\tau_A$. The determinant of the kinetic matrix is proportional to $(2c_3+c_4)^2\Tcal_0^2$; that is, in general there are three modes in the tensor perturbations around the cosmological background. However, the perturbation analysis around the Minkowski background showed that the spin-$2^-$ mode is a ghost mode. Therefore, in order to avoid a ghost or a strong coupling in the Minkowski limit we impose the degeneracy condition
\begin{align}
2c_3+c_4=0
\,, \label{deg_2^-}
\end{align}
which means that the $c_3,c_4$ terms must be given by the following form
\begin{align}
c_3 \epsilon^{\alpha\beta\rho\sigma}\Ra_{\mu\nu\rho\sigma}\Ta_{\alpha}{}^{\mu\nu}\Tcal_{\beta}
-2c_3 \Xcal^T_{\mu\nu}\Ta^{\mu\nu\rho}\Tcal_{\rho} 
=c_3 \epsilon^{\alpha\beta\rho\sigma}\CA_{\mu\nu\rho\sigma}\Ta_{\alpha}{}^{\mu\nu}\Tcal_{\beta}
\,.
\end{align}
After the field redefinition 
\begin{align}
\Xi_A=\tilde{\Xi}_A - 2(c_4+c_5) \epsilon_A \frac{\Tcal_0}{M_*^2}\tau_A
\,,
\end{align}
it becomes obvious that the variable $\tau_A$ is non-dynamical in the general cosmological background under the degeneracy condition \eqref{deg_2^-}. Removing $\tau_A$ by using its equation of motion from the Lagrangian, we obtain the quadratic Lagrangian in terms of the two variables $\tilde{\Xi}_A$ and $h_A$,
\begin{align}
S_T=\int  d\tau N a^3 \frac{M_{\rm pl}^2}{4} \sum_A\left[ 
\frac{1}{2} \frac{Y'{}^T}{N}\mathcal{K}\frac{Y'}{N}+Y^T\mathcal{M}\frac{Y'}{N}
-\frac{1}{2}Y^T\mathcal{V}Y 
\right]\,, \quad
Y=
\begin{pmatrix}
h_A \\
\tilde{\Xi}_A
\end{pmatrix}
\,,
\label{L_tensor}
\end{align}
where $\mathcal{K},\mathcal{M},\mathcal{V}$ are $2\times 2$ matrices.

We first consider the high $k$ limit. The asymptotic behaviour of the kinetic matrix is
\begin{align}
\mathcal{K}
\rightarrow \epsilon_A  \frac{ 2\Tcal_0 k}{M_*^2  a}
\begin{pmatrix}
c_3 & 0 \\
0 & 0
\end{pmatrix}
 +O(k^0)\,,
\end{align}
as $k\rightarrow \infty$. If $c_3\neq 0$ then one of the L or R modes is always a ghost in the high $k$ limit, similarly to the case in the presence of the gravitational Chern-Simons coupling~\cite{Dyda:2012rj}. We thus impose $c_3=c_4=0$ to prevent the ghost instability in the high $k$ limit hereinafter.

Under the conditions $c_3=c_4=0$, the kinetic matrix $\mathcal{K}$ has no $k$ dependency whereas $\mathcal{M}$ and $\mathcal{V}$ have the following $k$ dependency:
\begin{align}
\mathcal{M}&=\mathcal{M}^{(0)} + \epsilon_A \frac{ k}{a} \mathcal{M}^{(1)}
\,, \\
\mathcal{V}&= \mathcal{V}^{(0)} + \epsilon_A \frac{ k}{a} \mathcal{V}^{(1)} + \frac{k^2}{a^2}\mathcal{V}^{(2)}\,,
\end{align}
where $\mathcal{M}^{(i)}, \mathcal{V}^{(i)}$ are independent of $k$.
The components of the kinetic matrices are given by
\begin{align}
\mathcal{K}_{11}&=1-\alpha_{T_1} +\mathcal{K}_1 +\frac{c_5^2\Tcal_0^2}{\alpha_C M_*^2}  +\frac{\mathcal{K}_1}{\mathcal{K}_2} \Delta^2-2\Delta
\,, \\
\mathcal{K}_{12}&=\mathcal{K}_{21}=\frac{\mathcal{K}_1 \Delta}{\mathcal{K}_2}-1
\,, \\
\mathcal{K}_{22}&=\frac{\mathcal{K}_1}{\mathcal{K}_2}\,,
\end{align}
where we have defined
\begin{align}
\mathcal{K}_1&:=\bar{\lambda}-1+\alpha_{T_1}-\frac{3\bar{\Xi}_0+\bar{\Xi}}{6} -\left[ \frac{1}{2}(4c_1-c_2)+\frac{c_5^2}{\alpha_C} \right] \frac{{\Tcal}_0^2}{M_*^2}
\,,
\label{tensor_ghost_free}
\\
\mathcal{K}_2&:=\mathcal{K}_1\left(\mathcal{K}_1 +\frac{c_5^2\Tcal_0^2}{\alpha_C M_*^2}\right) +\bar{\varphi}^2\,,
\end{align}
and
\begin{align}
\Delta:=\frac{1}{6}(-3\bar{\Xi}_0+\bar{\Xi})+c_2\frac{\Tcal_0^2}{M_*^2}=\alpha_C(\hat{\rho}_B+\hat{p}_B)+c_2 \frac{\Tcal_0^2}{M_*^2}
\,.
\end{align}
The determinant of $\mathcal{K}$ is
\begin{align}
{\rm det}\mathcal{K}=\frac{ (1-\alpha_{T_1})\mathcal{K}_1-\bar{\varphi}^2 }{\mathcal{K}_2}
\,.
\end{align}
Therefore, the ghost free condition, namely the positive definiteness of $\mathcal{K}$, is reduced to 
\begin{align}
\mathcal{K}_1>\frac{\bar{\varphi}^2 }{1-\alpha_{T_1}}\,,
\label{gf_tensor}
\end{align}
under the stability conditions of the Minkowski spacetime \eqref{stability_M}. We note that $\mathcal{K}_1$ is necessary to be positive due the one of the stability condition \eqref{stability_M}, $0<\alpha_{T_1}<1$, and then $\mathcal{K}_2$ is also positive. The gradient terms are given by
\begin{align}
\mathcal{V}^{(2)}_{11}&=1-\alpha_{T_1} +\mathcal{K}_1 +\frac{c_5^2\Tcal_0^2}{\alpha_C M_*^2} 
\,, \\
\mathcal{V}^{(2)}_{12}&=\mathcal{V}^{(2)}_{21}=-1
\,, \\
\mathcal{V}^{(2)}_{22}&=\frac{1}{\mathcal{K}_2} \left( \mathcal{K}_1 +\frac{c_5^2\Tcal_0^2}{\alpha_C M_*^2} \right)
\,.
\end{align}
The determinant of $\mathcal{V}^{(2)}$,
\begin{align}
{\rm det}
\mathcal{V}^{(2)}
=\mathcal{K}_2^{-1}\left\{ \left[ (1-\alpha_{T_1})+\frac{c_5^2\Tcal_0^2}{\alpha_C M_*^2} \right] \left[ \mathcal{K}_1-\frac{\bar{\varphi}^2}{1-\alpha_{T_1}} + \frac{c_5^2\Tcal_0^2}{\alpha_C M_*^2}  \right]
+\frac{\bar{\varphi}^2}{1-\alpha_{T_1}} \frac{c_5^2\Tcal_0^2}{\alpha_C M_*^2}  \right\}
\,,
\end{align}
is positive under the stability condition \eqref{gf_tensor} as well as \eqref{stability_M}. Hence, the conditions \eqref{stability_M} and \eqref{gf_tensor} guarantee that the tensor perturbations has neither the ghost nor the gradient instability around the cosmological background.
The components of the matrices $\mathcal{M}^{(0)},\mathcal{M}^{(1)},\mathcal{V}^{(0)}$ and $\mathcal{V}^{(1)}$ are
\begin{align}
\mathcal{M}^{(0)}_{11}&=\mathcal{M}^{(0)}_{12}=\mathcal{M}^{(1)}_{11}=\mathcal{M}^{(1)}_{12}=0
\,, \\
\mathcal{M}^{(0)}_{21}&=\frac{2}{3}T_0+  \frac{\Delta }{\mathcal{K}_2}\left[ \mathcal{K}_1\left(H-\frac{1}{3}T_0 \right) 
-\frac{\alpha_C-2c_5}{2\alpha_C} \bar{\varphi} \Tcal_0 \right]
\,, \\
\mathcal{M}^{(0)}_{22}&=\frac{1}{\mathcal{K}_2}\left[ \mathcal{K}_1\left(H-\frac{1}{3}T_0 \right) 
-\frac{\alpha_C-2c_5}{2\alpha_C} \bar{\varphi} \Tcal_0 \right]
\,, \\
\mathcal{M}^{(1)}_{21}&=\frac{\bar{\varphi} \Delta }{\mathcal{K}_2} 
\,, \\
\mathcal{M}^{(1)}_{22}&= \frac{\bar{\varphi}}{\mathcal{K}_2}\,,
\end{align}
and
\begin{align}
\mathcal{V}^{(0)}_{11}&=\mathcal{V}^{(0)}_{12}=\mathcal{V}^{(0)}_{21}=
\mathcal{V}^{(1)}_{11}=\mathcal{V}^{(1)}_{12}=\mathcal{V}^{(1)}_{21}=0
\,, \\
\mathcal{V}^{(0)}_{22}&=\frac{M_*^2}{\alpha_C}-\frac{1}{\mathcal{K}_2}\left(H-\frac{1}{3}T_0 \right)\left[ \mathcal{K}_1 \left(H-\frac{1}{3}T_0 \right) - \frac{\alpha_C-2c_5}{\alpha_C} \bar{\varphi} \Tcal_0  \right]
\nn
&
+\frac{(\alpha_C-2c_5)^2\Tcal_0^2 }{4\alpha_C^2\mathcal{K}_2 } \left(\mathcal{K}_1 +\frac{c_5^2\Tcal_0^2}{\alpha_C M_*^2}  \right)
\,, \\
\mathcal{V}^{(1)}_{22}&=-2\frac{\bar{\varphi}} {\mathcal{K}_2}\left(H-\frac{1}{3}T_0 \right)
-\frac{(\alpha_C-2c_5)\Tcal_0 }{\alpha_C\mathcal{K}_2 } \left(\mathcal{K}_1 +\frac{c_5^2\Tcal_0^2}{\alpha_C M_*^2}  \right)\,,
\end{align}
where we retain the diagonal parts of $\mathcal{M}$ although they can be removed by taking integration by parts and by redefining $\mathcal{V}$.
The dispersion relation in the high $k$ limit is obtained by solving
\begin{align}
{\rm det}\left[ \omega^2\mathcal{K}+i\omega \frac{k}{a} (\mathcal{M}^{(1)}-\mathcal{M}^{(1)}{}^T)-\frac{k^2}{a^2} \mathcal{V}^{(2)} \right] \propto \left[\omega^4-(2+\delta_{\omega k})\omega^2\frac{k^2}{a^2} +(1+\delta_{kk})\frac{k^4}{a^4} \right]=0\,,
\end{align}
in terms of $\omega$.
The coefficients are given by
\begin{align}
\delta_{\omega k}&=\delta_{kk}+\frac{\Delta}{(1-\alpha_{T_1})\mathcal{K}_1-\bar{\varphi}^2} \left( \Delta -\frac{2c_5^2\Tcal_0^2}{\alpha_C M_*^2}  \right)
, \\
\delta_{kk}&=
\frac{c_5^2 \Tcal_0^2}{\alpha_C M_*^2[(1-\alpha_{T_1})\mathcal{K}_1-\bar{\varphi}^2]} 
\left[ (1-\alpha_{T_1})+\mathcal{K}_1 +\frac{c_5^2\Tcal_0^2}{\alpha_C M_*^2}   \right]\,.
\end{align}
The sound speed of the tensor perturbations is
\begin{align}
c_T^2=1+\frac{1}{2}\left( \delta_{\omega k}\pm \sqrt{\delta_{\omega k}^2+4(\delta_{\omega k}-\delta_{kk})} \right)
\,,
\end{align}
which deviates from unity in general.

The variable $h_A$, which represents the tensor perturbations of the metric, can be interpreted as the state to which matter fields directly couple. 
To derive the quadratic action of the tensor perturbations \eqref{L_tensor}, we have used the equations of motion for the tensor perturbations of the torsion $t_A,\tau_A$. Although we have not considered any matter field in the present paper, as far as the matter does not couple to $\Ta_{\mu\nu\rho}$, the equations of motion for $t_A$ and $\tau_A$ are unchanged even when a matter is introduced. In this case, anisotropic stress components of matter couple only with $h_A$ but does not with $\tilde{\Xi}_A$.
On the other hand, $h_A$ is not an eigenstate of the dispersion relation due to the existence of the non-diagonal components of matrices $\mathcal{K},\mathcal{M},\mathcal{V}$. We define variables $(\mathfrak{h}_A,\xi_A)$ via the relation
\begin{align}
h_A&= \mathfrak{h}_A - \mathcal{K}_{12} \xi_A
\\ 
\tilde{\Xi}_A & = \mathcal{K}_{11} \xi_A\,,
\end{align}
to diagonalize the kinetic matrix as well as the mass matrix $\mathcal{V}^{(0)}$. The set $(\mathfrak{h}_A,\xi_A)$ becomes the set of the mass eigenstates in the Minkowski limit $T_0,\mathcal{T}_0,\hat{\rho}_B,\hat{p}_B \rightarrow 0$ or in the torsionless de Sitter limit $T_0,\mathcal{T}_0,\Delta \rightarrow 0$ with $\hat{\rho}_B=-\hat{p}_B=$ constant. However, the gradient term $\mathcal{V}^{(2)}$ and the friction terms $\mathcal{M}$ have non-diagonal components in the general cosmological background and then two modes cannot be decoupled.

Therefore, the tensor perturbations in the generic case have qualitatively different features from those in the single field limit discussed in Sec.~\ref{sec_heavy}. Gravitational waves are no longer freely propagating massless waves with the speed of light in general cosmological background. Furthermore, the L and R modes obey different equations when the background universe has a non-zero value of $\Tcal_0$. The parity invariance is broken spontaneously. It must be intriguing to study observational effects of these features. However, we leave them for a future study since we also have to study scalar (and vector) perturbations to discuss observational constraints on the model which are beyond the scope of the present paper.

Before closing this section, we discuss the heavy mass limit of the spin-$2^+$ particle.
Let us consider the limit $\alpha_C \rightarrow 0$ in order to have the infinite mass limit of the spin-$2^+$ particle where the degree of freedom of $2^+$ may be integrated out. To satisfy the ghost-free condition \eqref{gf_tensor}, the coupling constant $c_5$ has to vanish as well under the limit $\alpha_C \rightarrow 0$. Hence, we take the limit $\alpha_C,c_5 \rightarrow 0 $ with $c_5^2/\alpha_C< \infty$ where the divergence of $\mathcal{V}^{(2)}_{11}$ and $\mathcal{M}^{(0)}_{21}$ can be compensated by taking the normalization $\tilde{\Xi}_A \rightarrow \alpha_C^{1/2}\tilde{\Xi}_A$. After integrating out $\tilde{\Xi}_A$ under the limit, we obtain the quadratic Lagrangian for the massless tensor mode,
\begin{align}
S_T|_{\alpha_C \rightarrow 0}=
\int d\tau N a^3 \frac{M_{\rm pl}^2 }{8} \frac{\mathcal{F}_T}{c_T^2} \left[\left(\frac{h_A'}{N}\right)^2 - \frac{c_T^2 k^2}{a^2} h_A^2 \right]
\,,
\label{tensor_massless}
\end{align}
where
\begin{align}
\mathcal{F}_T&=\bar{\lambda} +(c_2-4c_1) \frac{\Tcal_0^2}{2M_*^2}
\,, \\
c_T^2&=\left[ 1-\frac{c_2 \mathcal{T}_0^2 }{\mathcal{F}_T M_*^2}\left( 2 + \frac{c_2 \Tcal_0^2}{M_*^2} \frac{1-\alpha_{T_1}-\mathcal{F}_T} {(1-\alpha_{T_1}-\mathcal{F}_T)^2+\bar{\varphi}^2} \right) \right]^{-1}\,.
\end{align}
From this expression one can straightforwardly compute the tensor power spectrum and the spectral tilt, following e.g.~\cite{Kobayashi:2011nu}.
The parity invariance is recovered in the infinitely heavy mass limit of the spin-$2^+$ particle while the speed of gravitational waves is still different from the speed of light due to the coupling $c_2 \RA^T_{\mu\nu}\Tcal^{\mu}\Tcal^{\nu}$.
The result of the single field limit is obtained when we furthermore consider the background with $\Tcal_0 \rightarrow 0$. This implies that the evolution of tensor perturbations during and after inflation is generally the same as that of the single field limit when $\alpha_C\rightarrow 0$ and $m_{{\rm eff},\theta}^2>0$ since $\Tcal_0$ can go to zero before $N_e=50$.

\section{Summary}
\label{summary}
The idea that gravity is interpreted as a gauge force predicts the existence of additional massive particle species carrying the gravitational interactions which can be seen in high energy phenomena. The present paper has studied the inflationary background dynamics of the universe and the linear tensor perturbations in a ghost-free quadratic gravity with a dynamical torsion. The underlying spacetime geometry is the Riemann-Cartan geometry (or the Weyl-Cartan geometry as a result of the invariance under a projective transformation), where the torsion as well as the curvature play central roles.

We first formulate the Lagrangian so that the action asymptotically has the local Weyl invariance in the UV limit. In particular, we have focused on a theory which consists of the massive spin-$2^+,1^+,0^+,0^-$ particle species in addition to the massless graviton, where the number and $\pm$ represents the spin and the parity, respectively. This model reproduces the Starobinsky model under the limit where the $2^+,1^+,0^-$ particles become infinitely heavy with the mass of $0^+$ kept finite. The spectral index and the tensor-to-scalar ratio are shown in Fig.~\ref{fig_ns_r} under the single field limit. This result should be viewed as a reference value since in general the $2^+,1^+,0^-$ must have finite masses and then contribute the observables, more or less. We then study the model in the generic parameter space. As for the background dynamics in which only $0^+$ and $0^-$ are dynamical, we find an useful field transformation \eqref{change}-\eqref{conformal_A} by which we obtain the quasi-Einstein frame for the minisuperspace action. In the quasi-Einstein frame the model is characterized by a two-dimensional hyperbolic field space and a field potential which is a combination of those of a Starobinsky-like inflation and a natural inflation. The qualitative behaviour of the inflationary dynamics is determined by the sign of the effective mass squared of the $0^-$ in the high energy limit denoted by $m_{\theta, {\rm eff}}^2$. The overall behavior of the background dynamics is shown in Figs.~\ref{fig_PS_alphaC} and \ref{fig_PS_g3}. When $m_{\theta, {\rm eff}}^2>0$, the background value of $0^-$ decays and then the Starobinsky-like inflationary universe is obtained even if the $0^+$ is not at the top of the potential initially. This would be a remarkable feature because we do not need a fine-tuning of the initial condition for the inflaton. On the other hand, in the case of $m_{\theta, {\rm eff} }^2\lesssim 0$, the $0^-$ has a non-zero background value and other inflationary scenarios can be obtained. From the tensor perturbation analysis, we have learned that the gravitational wave is no longer a freely propagating massless wave around the generic cosmological background and that the parity invariance is spontaneously broken due the background value of the $0^-$.

Hence, it should be interesting to perform detailed studies on observables, combining the results of the present paper with analysis on scalar (and vector) perturbations, at linear order and non-linear orders. In particular, if the future observations detect the primordial gravitational waves at the level of $r\sim 10^{-3}$ consistently with the $R^2$ models of inflation, studies on generic quadratic gravity with the dynamical torsion (and the non-metricity) will be obviously important. Inflation can then be used to reveal the underlying nature of gravity. We leave further analysis on inflation for future studies.

In the present paper, we have assumed either that the local Weyl invariance is restored in the UV or that it is an approximate symmetry in the intermediate scales $M_* \lesssim E \ll \Lambda$. As argued in section~\ref{sec_QG}, this assumption may be justified if the renormalization group (RG) flow of the underlining theory admits a UV fixed point with the local Weyl invariance or a saddle point with the local Weyl invariance. It is certainly important to investigate the RG flow of concrete theories to see if this is the case.

\section*{Acknowledgments}
K.A. acknowledges the xTras package~\cite{Nutma:2013zea} which was used for tensorial calculations.
The work of K.A. was supported in part by Grants-in-Aid from the Scientific Research Fund of the Japan Society for the Promotion of Science  (No.~19J00895). The work of S.M. was supported by Japan Society for the Promotion of Science Grants-in-Aid for Scientific Research No.~17H02890, No.~17H06359, and by World Premier International Research Center Initiative, MEXT, Japan. 


\appendix
\section{Einstein frame in weakly curved spacetime}
\label{appendix}
In this section, we show that the Lagrangian \eqref{LG} with \eqref{GF_L2}, \eqref{GF_L4}, \eqref{L_4^3} and \eqref{L_4^4} indeed has non-ghost massive spin-$2^+,1^+,0^+,0^-$ particle species when the curvature and the torsion are small, namely 
\begin{align}
| \RA_{\mu\nu\rho\sigma}| \ll M_*^{2}
\,, \quad |  T_{\mu\nu\rho}| \ll M_*
\,. \label{weak_curve}
\end{align}
For simplicity, we assume that matter fields directly couple with the metric only. 

By the use of auxiliary variables $\lambda,\varphi,\Xi_{\mu\nu},A_{\mu\nu}$, where $\Xi_{\mu\nu}=\Xi_{(\mu\nu)}$ and $A_{\mu\nu}=A_{[\mu\nu]}$ are symmetric and antisymmetric tensors, respectively, we obtain the equivalent Lagrangian \eqref{L_eq} with a matter action, $S_{\rm m}=\int d^4x \sqrt{-g}\mathcal{L}_{\rm m}$. The original Lagrangian is obtained when integrating out all auxiliary variables $\lambda,\varphi,\Xi_{\mu\nu},A_{\mu\nu}$ which are solved as
\begin{align}
\lambda-1&=\frac{\alpha_R}{3M_*^2}\RA
\,, \\
\varphi&=\frac{\alpha_{\Xcal} }{3M_*^2}\Xcal
\,, \\
\Xi_{\mu\nu}&=\frac{2\alpha_C}{M_*^2}\left( \RA_{(\mu\nu)}-\frac{1}{6}g_{\mu\nu}\RA \right)
\,, \\
A_{\mu\nu}&=\frac{6\alpha_Y}{M_*^2}Y_{\mu\nu}
\,,
\end{align}
respectively.

The assumption \eqref{weak_curve} concludes 
\begin{align}
|\delta\lambda|\,,\   |\Xi_{\mu\nu}|\,,\   |\varphi|\,,\  |A_{\mu\nu}| \ll 1 \,, 
\end{align}
where $\delta\lambda := \lambda-1 $. 
We take the transformation
\begin{align}
\Xi_{\mu\nu} &=\xi_{\mu\nu}+\frac{4\alpha_C(1-\alpha)}{3M_*^2 \alpha} \nabla_{(\mu}T_{\nu)} +\frac{2\alpha_C}{M_*^2} \frac{\alpha_{T_1}-\alpha}{\alpha_{T_1}}\nabla_{\mu}\nabla_{\nu}\lambda\,,
\end{align}
and then take the variation with respect to the torsion. The equation of motion of the torsion leads to
\begin{align}
T_{\mu\nu\rho}&=\frac{1}{\alpha_{T_2}}g_{\mu[\nu}\nabla_{\rho} \delta\lambda+\frac{1}{\alpha_{T_3} }\epsilon_{\mu\nu\rho\sigma}\nabla^{\sigma}\varphi
+\frac{1}{\alpha_{T_1}}\nabla_{[\nu}\xi_{\rho]\mu}
\nn
&-\frac{1}{3}\left[ \left(\frac{1}{\alpha_{T_1}}-\frac{1}{\alpha_{T_2}}\right)\nabla_{\mu}A_{\nu\rho}-\left(\frac{1}{\alpha_{T_1}}+\frac{2}{\alpha_{T_3}}\right)\nabla_{[\nu}A_{\rho]\mu} \right]
\nn
&+\frac{1}{3}\left( \frac{1}{\alpha_{T_1}}-\frac{1}{\alpha \alpha_{T_2}} \right)
\left(g_{\mu[\nu}\nabla^{\sigma}\xi_{\rho]\sigma}-g_{\mu[\nu}\nabla_{\rho}\xi^{\sigma}{}_{\sigma}+g_{\mu[\nu}\nabla^{\sigma}A_{\rho]\sigma} \right)
+\cdots\,,
\label{solT}
\end{align}
where $\cdots$ represents terms which can be ignored under the assumption \eqref{weak_curve}. We can thus integrate out the torsion as far as \eqref{weak_curve} is satisfied. After substituting \eqref{solT} into \eqref{L_eq}, we obtain
\begin{align}
\mathcal{L}_{\rm eq}&=\frac{M_{\rm pl}^2}{2} \bigg[ 
(1+ \delta \lambda) R(g) +\xi^{\mu\nu} G_{\mu\nu} +\frac{1}{2\alpha_{T_1}}\mathcal{L}_{\rm EH}^{(2)}(\xi)-\frac{M_*^2}{4\alpha_C} (\xi_{\mu\nu}\xi^{\mu\nu}-\xi^{\mu}{}_{\mu}\xi^{\nu}{}_{\nu} )
\nn
& \qquad \qquad
+\frac{3\alpha^2 }{2\alpha_{T_1}} \nabla_{\mu} \delta\lambda \nabla^{\mu}\delta\lambda -\frac{3M_*^2}{2\alpha_R}\delta\lambda^2
+ \nabla_{\mu} \delta \lambda (\nabla_{\nu}\xi^{\mu\nu} - \nabla^{\mu} \xi^{\nu}{}_{\nu} )
\nn
& \qquad \qquad
-\frac{1}{36 }\left(\frac{1}{\alpha_{T_1}}-\frac{1}{\alpha_{T_3}}\right) F_{\mu\nu\rho}F^{\mu\nu\rho} -\frac{M_*^2}{12 \alpha_Y} A_{\mu\nu}A^{\mu\nu}
-\frac{3}{2\alpha_{T_3}} \nabla_{\mu} \varphi \nabla^{\mu}\varphi -\frac{3M_*^2}{2\alpha_{\Xcal}} \varphi^2
+\cdots
 \bigg]
 \nn
&+ \mathcal{L}_{\rm m}(g,\psi)
\,,
\end{align}
where
\begin{align}
\mathcal{L}_{\rm EH}^{(2)}(\xi)&=-\frac{1}{2}\nabla_{\rho}\xi_{\mu\nu}\nabla^{\rho}\xi^{\mu\nu}+\nabla_{\rho}\xi_{\mu\nu}\nabla^{\nu}\xi^{\mu\rho}-\nabla_{\mu}\xi^{\rho}{}_{\rho}\nabla_{\nu}\xi^{\mu\nu}+\frac{1}{2}\nabla_{\mu}\xi^{\nu}{}_{\nu}\nabla^{\mu}\xi^{\rho}{}_{\rho}
\,,
\\
F_{\mu\nu\rho} &=3\partial_{[\mu } A_{\nu\rho]}
\,,
\end{align}
and we have used $\alpha_{T_2}=\alpha_{T_1}/\alpha^2$. We can then move to the Einstein frame $g^E_{\mu\nu}$ via
\begin{align}
g_{\mu\nu}=\lambda^{-1} g^E_{\mu\nu} +\xi_{\mu\nu}\,.
\end{align}
The action in the Einstein frame is given by
\begin{align}
S=\int d^4x \sqrt{-g^E} \frac{M_{\rm pl}^2}{2} \Biggl[ &
R(g^E) +\frac{1}{2} \left(\frac{1}{\alpha_{T_1}}-1 \right) \mathcal{L}_{\rm EH}^{(2)}(\xi) - \frac{M_*^2}{4\alpha_C} (\xi_{\mu\nu}\xi^{\mu\nu}-\xi^{\mu}{}_{\mu}\xi^{\nu}{}_{\nu} )
\nn
&
-\frac{3}{2}\left(1-\frac{\alpha^2}{\alpha_{T_1}} \right) (\nabla \delta \lambda)^2 -\frac{3 M_*^2}{2\alpha_R} \delta \lambda^2
\nn
&
-\frac{1}{36 }\left( \frac{1}{\alpha_{T_1} }-\frac{1}{\alpha_{T_3} } \right) F_{\mu\nu\rho}F^{\mu\nu\rho} -\frac{M_*^2}{12 \alpha_Y} A_{\mu\nu}A^{\mu\nu}
\nn
&
-\frac{3}{2\alpha_{T_3}} (\nabla \varphi)^2 -\frac{3M_*^2}{2\alpha_{\Xcal}} \varphi^2 +\cdots \Biggl]
+S_{\rm m}
\,.
\end{align}
In this expression, the covariant derivatives and the contractions of the indices are computed by the original metric $g_{\mu\nu}$; however, these quantities can be replaced with those computed by the Einstein frame metric $g^E_{\mu\nu}$ without any change of the expression under the approximation \eqref{weak_curve} at the leading order.
In the Einstein frame, the non-ghost massive spin-$2^+,1^+,0^+,0^-$ particles are represented by the fields $\xi_{\mu\nu},A_{\mu\nu}, \delta\lambda, \varphi$, respectively.

\bibliography{ref}
\bibliographystyle{JHEP}

\end{document}